%% file: arxiv.tex
\documentclass[final,authoryear,5p,times, twocolumn]{article}

\usepackage{lineno}
\usepackage{graphicx}
\usepackage{color}
\usepackage[dvipsnames]{xcolor}
\usepackage{amssymb}
\usepackage{amsmath}
\usepackage{bigints}
\usepackage[bookmarksdepth=3]{hyperref}
\usepackage{lineno}
\usepackage{mathtools}
\usepackage{blkarray}
\usepackage[capposition=bottom]{floatrow}
\usepackage{varwidth}
\usepackage{pifont}
\graphicspath{{}}
\usepackage{booktabs}
\usepackage{pgf,tikz}
\usepackage{mathrsfs}
\usepackage{wrapfig}
\usepackage{authblk}
\usetikzlibrary{arrows}
\hypersetup{
    colorlinks=true,
    citecolor=red,
    linkcolor=blue,
    filecolor=blue,      
    urlcolor=red,}

\hypersetup{colorlinks=True, citecolor=NavyBlue, linkcolor=NavyBlue,
urlcolor=NavyBlue}

\usepackage[letterpaper , left=.5 in, right=.5 in, top= 1 in, bottom=1 in, marginparwidth=0 cm, marginparsep=0cm]{geometry}

\usepackage[ruled,vlined]{algorithm2e}

\title{Algebraic 3D Graphic Statics: Constrained Areas}

\author[$a$]{Masoud Akbarzadeh}

\author[$a, b$]{M\'{a}rton Hablicsek}

\affil[$a$]{Polyhedral Structures Laboratory, School of Design, University of Pennsylvania, Philadelphia, USA}

\affil[$b$]{Department of Mathematics, Leiden University, Leiden, Netherlands}

\begin{document}


\twocolumn[
  \begin{@twocolumnfalse}
    \maketitle
\begin{abstract}
This research is a continuation of the Algebraic 3D Graphic Statics Methods. It provides algorithms and (numerical) methods to geometrically control the magnitude of the internal and external forces in the reciprocal diagrams of 3D/Polyhedral Graphic statics. 3D graphic statics (3DGS) is a recently rediscovered method of structural form-finding based on a 150-year old proposition by Rankine and Maxwell in Philosophical Magazine. In 3DGS, the form of the structure and its equilibrium of forces is represented by two polyhedral diagrams that are geometrically and topologically related. The areas of the faces of the force diagram represent the magnitude of the internal and external forces in the system. For the first time, the methods of this research allow the user to control and constrain the areas and edge lengths of the faces of general polyhedrons that can be convex, self-intersecting, or concave. As a result, a designer can explicitly control the force magnitudes in the force diagram and explore the equilibrium of a variety of compression and tension-combined funicular structural forms. In this method, a quadratic formulation is used to compute the area of a single face based on its edge lengths. The approach is applied to manipulating the face geometry with a predefined area and the edge lengths. Subsequently, the geometry of the polyhedron is updated with newly changed faces. This approach is a multi-step algorithm where each step includes computing the geometry of a single face and updating the polyhedral geometry. One of the unique results of this framework is the construction of the zero-area, self-intersecting faces, where the sum of the signed areas of a self-intersecting face is zero, representing a member with zero force in the form diagram. The methodology of this research can clarify the equilibrium of some systems that could not be previously justified using reciprocal polyhedral diagrams. Therefore, it generalizes the principle of the equilibrium of polyhedral frames and opens a completely new horizon in the design of highly-sophisticated funicular polyhedral structures that are beyond compression-only systems.\\


\vspace{0.5cm}

\end{abstract}
  \end{@twocolumnfalse}
]



 \input{introduction.tex}
 \input{nomen}

\input{methodology.tex}

\input{implementation}

\input{application.tex}
\input{conclusion.tex}

\bibliographystyle{plain}
\bibliography{3DGS_refs}

\end{document}

%% file: introduction.tex
\section{Introduction}

Recently, the geometry-based structural design methods, known as Graphic Statics, has been extended to 3D dimensions based on a historic proposition by Rankine and Maxwell in Philosophical Magazine \cite{Rankine1864b, Akbarzadeh2013,Akbarzadeh2016_phd,Beghini2013a, McRobie2016, maxwell1864, Lee2018phd, Lee2018, McRobie2016, McRobie2017a, McRobie2017b}.




\begin{figure}
    \centering
    \includegraphics[width = 0.9\columnwidth]{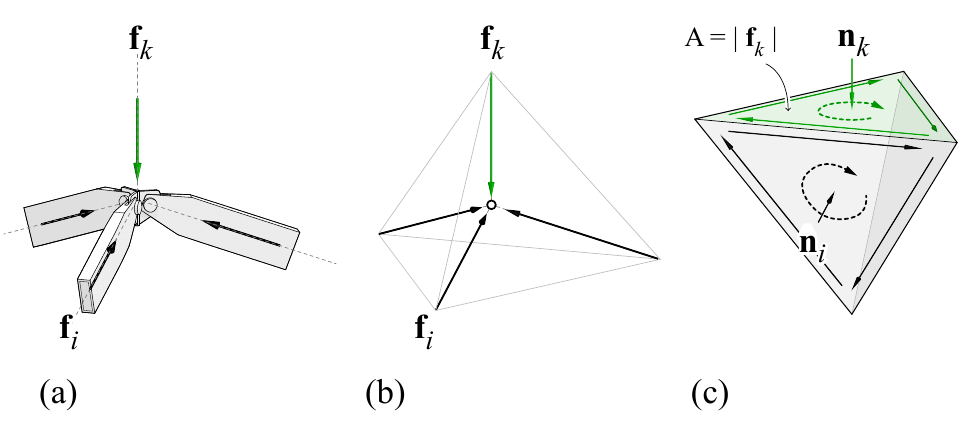}
    \caption{(a) A 3D structural joint with an applied force and internal forces in its members; (b) the form diagram/bar-node representation of the same joint in the context of 3DGS; and (c) the force diagram/polyhedron representing the equilibrium of the same node in 3DGS.}
    \label{fig:3d}
\end{figure}

\begin{figure}
    \centering
    \includegraphics[width = \columnwidth]{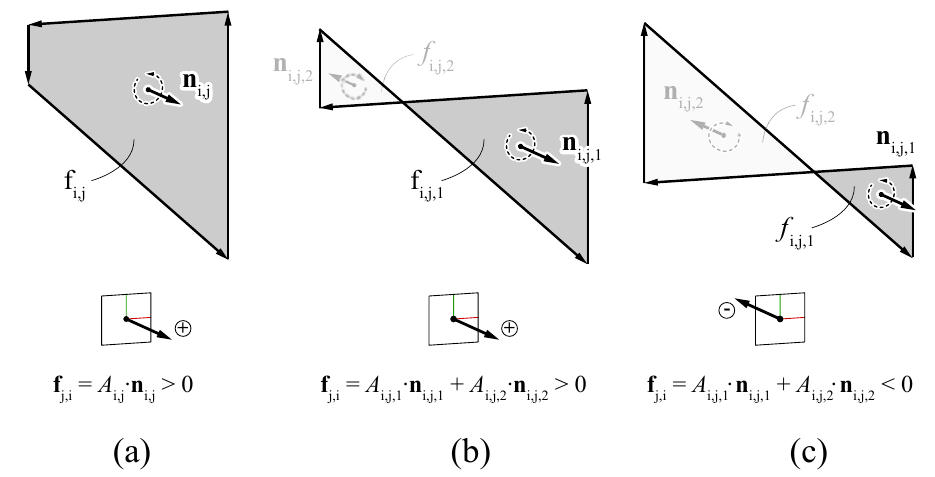}
    \caption{From left to right: (a) a convex face with a positive force direction (out of the page); (b) a complex face with two enclosed regions and a positive net force direction; (c) and a complex face with two enclosed regions and a negative net force direction.}
    \label{fig:indeterminate}
\end{figure}

In this method which is called \textit{3D Graphical Statics using Reciprocal Polyhedral Diagrams}, the equilibrium of the forces in a single node is represented by a closed \textit{polyhedron} or a polyhedral \textit{cell} with planar faces (Figure \ref{fig:3d}a). Each face of the force polyhedron is perpendicular to an edge in the form diagram, and the magnitude of the force in the corresponding edge is equal to the area of the face in the force polyhedron. The sum of all area-weighted normals of the cell must equal zero that can be proved using the divergence theorem \cite{Stokes1905, akbarzadeh2015_convex,Akbarzadeh2015_3dsub, Akbarzadeh2015_global,Akbarzadeh2016_phd,McRobie2016}. In some cases, a cell can have \textit{complex faces} (self-intersecting) and which has multiple enclosed regions (Figure \ref{fig:indeterminate}a). The direction and the magnitude of the force corresponding to a complex face can be determined by summing the area-weighted normals of all of the enclosed regions (Fig. \ref{fig:indeterminate}b, c). As a result the direction of the internal force in the members of the structure might flip based on the direction of the face of a single force cell.

Exploiting the potentials of 3DGS in design and engineering requires the ability to manipulate the geometric diagrams without breaking the reciprocity between the two and instantly observe the effect of the change in the other diagram. The existing computational tools for the design and manipulation of the reciprocal polyhedrons of 3DGS are quite limited. Moreover, controlling and optimizing the magnitude of forces by changing the areas of the faces needs efficient algorithms.




\subsection{Related works}

In 2016, \cite{Akbarzadeh2016_phd} showed that the reciprocal diagrams of 3DGS can be constructed in a procedural (step-by-step) approach in a parametric software by assigning constraints between reciprocal components of each diagram that allows simultaneous control over the geometry of both diagrams. This method is extremely time-consuming and tedious for structures with a large number of nodes and members \cite{Akbarzadeh2015_global, Akbarzadeh2016_phd}. \cite{akbarzadeh2015_convex} developed a computational algorithm that could receive convex polyhedral cells as a primal and construct its reciprocal diagram iteratively within a certain tolerance defined by the user \cite{akbarzadeh2015_convex, Akbarzadeh2016_phd}. This method cannot deal with (non-convex) self-intersecting polyhedrons or explore tension and compression equilibrium. Moreover, controlling the areas of the faces was computationally quite expensive. In 2018, \cite{Lee2018} proposed a method called \textit{Disjointed Force Polyhedra} where the equilibrium of the system was computed by constructing a single convex polyhedron for each node using Extended Gaussian Image algorithm \cite{Little1983, Horn1984, Moni1990} and matching the areas of the shared faces \cite{Lee2017, Lee2018}. This method allows controlling the areas of the convex cells, but it breaks the reciprocity between the two diagrams. Moreover, it cannot control the areas of the self-intersecting faces. Recently, \cite{HABLICSEK201930} developed an algebraic formulation relating the geometry of the reciprocal polyhedral diagrams using a linear system of equations. This method can directly construct the dual from a given primal in one step \cite{akbarzadeh2018developing, HABLICSEK201930, akbarzadeh2020geometric}. Although the previous formulation could immediately construct the reciprocal polyhedral diagrams, it did not provide any insight on how to control the areas of the faces corresponding to the magnitude of the forces in the form diagram. Moreover, the geometrically constrained constructions were also not addressed.

\subsection{Contributions}
This paper provides a robust algebraic method to construct polyhedrons with assigned areas and edge lengths of their faces from which its reciprocal dual can be constructed as the structural form. The formulation introduced in this paper relates the areas of the faces of the polyhedral system to its edge lengths allowing the combination of this method with the previous algebraic formulation to control the areas of the faces. Moreover, the methods of this research can compute the areas of self-intersecting faces with constraint edges which has never been addressed in the literate previously. Specifically, this approach can construct zero-area, self-intersecting faces in the system, where the sum of the signed areas of a self-intersecting face is zero. The existence of such faces in the force diagram allows to either remove forces in the boundary conditions or internal forces and therefore, describe internal force equilibrium that previously was not possible using reciprocal polyhedral diagrams.

The paper is organized as follows. A quadratic formulation to compute the area is introduced for a single face based on its edge lengths (Section \ref{sec:area}). Then, a methodology is described to manipulate the geometry of the face with a predefined area and edge lengths (Section \ref{sec:face}). Subsequently, the geometry of the polyhedron is updated with the newly changed faces (Section \ref{sec:poly}). This approach is a multi-step algorithm where each step includes computing the geometry of a single face and updating the polyhedral geometry. In the end, the dual structural form is updated with the new magnitude of the internal or external forces (Section \ref{sec:dual}). Section \ref{sec:app} shows the application of this method in the design of funicular structures with zero force members or reactions in the boundary conditions.



%% file: nomen.tex
\subsection{Nomenclature}

We denote the algebra objects of this paper as follows; matrices are denoted by bold capital letters (e.g. $\textbf{A}$); vectors are denoted by lowercase, bold letters (e.g., $\textbf{v}$), except the user input vectors which are represented by the Greek letters ($\nu$ and $\xi$). Table \ref{table:tab2} encompasses all the notation used in the paper.

\begin{table}
\caption{Nomenclature for the symbols used in this paper and their corresponding descriptions.}
\resizebox{\columnwidth}{!}{%
\centering
\begin{tabular}{ll}
\hline

 \textit{Topology}       & \textit{Description}                                                                                       \\ \hline
 $\Gamma$                        & primal diagram                                                                                    \\
 $\Gamma^\dagger$                & dual, reciprocal diagram                                                                          \\
 $v$                             & \# of vertices of $\Gamma$                                                                    \\
 $e$                             & \# of edges of $\Gamma$                                                                       \\
 $f$                             & \# of faces of $\Gamma$                                                                       \\
 $c$                             & \# of cells of $\Gamma$                                                                       \\
 $v^\dagger$                     & \# of vertices of $\Gamma^\dagger$                                                            \\
 $e^\dagger$                     & \# of edges of $\Gamma^\dagger$                                                               \\
 $f^\dagger$                     & \# of faces of $\Gamma^\dagger$                                                               \\
\hline
\textit{Matrices}               &                                                       \\ \hline    
$\mathbf{M}_f$ &  Area matrix of the face $f$\\

$\mathbf{E}_f$ &  Equilibrium matrix of the face $f$\\

$\mathbf{L}_f$ & Matrix of predefined edge lengths of the face $f$\\

$\mathbf{E}_p$ &  Equilibrium matrix of the polyhedral system $p$\\

$\mathbf{E}^\dagger$ &  Equilibrium matrix of the dual\\

$\mathbf{L}_p$ & Matrix of predefined edge lengths of the polyhedron $p$\\

$\mathbf{B}_f$ & Constraint matrix of the face $f$\\

$\mathbf{B}_p$ &  Constraint matrix of the polyhedron $p$\\

$\mathbf{B}_p^+$ &  Moore-Penrose inverse of $\mathbf{B}_p$\\

$\textbf{RREF}_f$ & RREF of $\left(textbf{B}_f|\textbf{b}_f\right)$\\

 $\textbf{E}^\dagger$                    & equilibrium matrix of the dual diagram                                                                                \\

 $(\textbf{E}^{\dagger})^+$              & Moore-Penrose inverse of $\textbf{E}^\dagger$                                                             \\
\hline
 \textit{Vectors}                &                                                       \\ \hline   
 
$\mathbf{n}$ & consistent unit normal vector \\

 $\mathbf{q}$ &  vector of edge lengths\\

$\mathbf{u}_j$ & direction vector of edge vector $\mathbf{e}_j$\\

$\mathbf{b}_f$ &  constraint vector for the face $f$\\

$\mathbf{l}_f$ & vector of predefined edge lengths of the face $f$\\

$\mathbf{b}_p$ &  constraint vector for the polyhedron $p$\\

$\mathbf{l}_p$ & vector of predefined edge lengths of the polyhedron $p$\\

$\mathbf{q}_{nci}$ &  vector of nci edge lengths of a face $f_i$\\

$\mathbf{q}_{fix}$ &  vector of fixed edge lengths of a face $f_i$\\

$\mathbf{q}_{nfd}$ &  vector of nfd edge lengths of a face $f_i$\\

$\mathbf{q}_{ci}$ &  vector corresponding to the edge length of the ci edge\\

$\mathbf{e}_j^\dagger$ & edge vector of $e_j^\dagger$ in $\Gamma^\dagger$\\

$\mathbf{u}_j^\dagger$ & direction vector of edge vector $\mathbf{e}_j^\dagger$\\

 $\textbf{q}^\dagger$                    & vector of edge lengths of $\Gamma^\dagger$                     \\
\hline
\textit{Parameters}          &                                                         \\ \hline    
$\nu$                           & parameter for the MPI method to solve Eq. \ref{eq:mpi}\\
 $\xi$                         & parameter for the MPI method to solve Eq. \ref{eq:04}                                        \\
\hline

\textit{Other}  &\\
\hline
$O$ & centroid of a face\\

$h_{i,j}$ & (signed) distance of $v_j$ from $e_i$\\

$H_i$ & average (signed) distance of the $v_j$ from $e_i$\\

$A_f$ & area of face $f$\\

$\mu_{i,j}$ & the scalar ratio between $\textbf{e}_i\times \textbf{e}_j$ and $\textbf{n}$\\

$\eta_{i,j}$ & the scalar ratio between $\textbf{u}_i\times \textbf{u}_j$ and $\textbf{n}$\\

$GDoF_f$ & Geometric Degrees of Freedom of face $f$\\

$CGDoF_f$ & Constrained Geometric Degrees of Freedom of face $f$\\ 

$e_{fix}$ & (user)-selected fixed edges of a face\\

$e_{ind}$ & list of independent edges of a face\\

$e_{nfd}$ & list of nfd edges of a face\\

$e_{nci}$ & list of nci edges of a face\\

$e_{fix}^p$ & (user-)selected fixed edges of the polyhedron\\

$q_{ci}$ & length of the ci edge\\

$r$    & rank of $\textbf{RREF}_f$       \\
\hline  
\end{tabular}
}
\label{table:tab2}
\end{table}

%% file: methodology.tex
\section{Research Methods}\label{sec:meth}
\subsection{Overview}

The first step in the methodology is to link the areas of the faces of the polyhedral system to their edge lengths. The definition of the area based on the edge lengths will result in a quadratic function per face of the polyhedron. As a result, controlling the areas of the faces of the polyhedral system requires solving a complex system of non-homogeneous quadratic equations simultaneously. This complex system of quadratic equations is usually solved using optimization methods. However, to the knowledge of the authors, these methods usually fail in computing self-intersecting/compression- and-tension combined systems. Moreover, the objective of our research is to provide a methodology to control the areas of the faces without perturbing the system drastically. Therefore, in this section, we provide a simple methodology to solve these quadratic equations sequentially by preserving the main geometric features of both the form and force diagrams.  

In this method, there are two types of equations; (i) the quadratic equations that compute the areas of the faces based on the edge lengths; and (ii) the linear equations that provide the geometry of the faces of the polyhedrons with user-defined edge lengths as constraints. The quadratic equations of the faces are solved using the linear equations around the edges of each face with constrained edge lengths. Each quadratic equation for a face area has as many variables as the number of edges of the face which results in a variety of significantly different solutions. However, we can control the solution space by reducing the number of variables to one. This allows us to find a solution for the quadratic equation with a limited geometric perturbation in the system. 

In the following sections, we introduce the steps to develop the quadratic equation to compute the area of a face of a polyhedral system based on the edge lengths, and then we develop the non-homogeneous linear equation system describing the equilibrium equation for the face with predefined edge lengths. In the end, we show how to solve the equation system and how to recompute the geometry of the form.

\input{method_01.tex}

\input{method_02.tex}

\input{method_03.tex}
\input{method_04.tex}

\input{method_05.tex}

%% file: method_01.tex





\subsection{Linear equilibrium equations for a polyhedral system}

 In the previous paper \cite{HABLICSEK201930}, we showed how to write the equilibrium equations for a system of polyhedral cells with planar faces. For each face $f_i$, we can write an equation based on its edge lengths that shows the closeness of the face. By choosing a normal vector for each face $f_i$, we can obtain a consistent orientation of the edges. We denote the unit direction vector $\textbf{u}_{j}$ corresponding to the edge vector $\textbf{e}_{j}$ of Figure \ref{fig:area}a. Since each face provides a closed loop of edges, the sum of the edge vectors has to be the zero vector. Thus, we obtain a vector equation for the edge lengths $q_j$ of $\mathbf{e}_j$ as
\begin{equation}\label{eq:eqsysface}
    \sum_{e_{j}}\textbf{u}_{j}q_j=\textbf{0}
\end{equation}
where the sum runs over the edges $e_{j}$ of the face $f_i$.
 
 Thus, each face $f_i$ of the polyhedron $p$ provides three equations for the edge lengths, which can be described by a $[3\times e]$ matrix, $\mathbf{E}_{f_i}$
 \begin{equation}\label{eq:eqface}
     \mathbf{E}_{f_{i}}\mathbf{q}=\mathbf{0}
 \end{equation}
 where $\mathbf{q}$ denotes the vector of the edge lengths of the polyhedron. This equation describes the geometry of the face.
 
 Similarly, we can obtain a $[3f \times e]$ matrix, $\textbf{E}_p$ describing the geometry of the entire network. Here $f$ denotes the number of faces and $e$ denotes the number of edges in the network. In other words, we have a linear equation system
\begin{equation}\label{eq:eqpoly}
    \textbf{E}_p\textbf{q}=\textbf{0}
\end{equation}
where $\textbf{q}$ denotes (again) the vector of edge lengths of the polyhedron $p$. Each solution of the linear Eq. system \ref{eq:eqpoly} represents a network, whose edges are parallel to their associated edges of the original network with different edge lengths.




%% file: method_02.tex
\subsection{Quadratic equation system for the area of a face}
\label{sec:area}
In this section, we explain, how to develop a quadratic system of equation for a face $f_i$ of a polyhedral network based on the edge vectors of the face after \cite{ORourke2011}. 


 Consider a face $f_i$ with $k$ vertices: $v_0$, $v_1$, ..., $v_{k-1}$. We denote the edge $e_i$ by the vertices $v_i$ and $v_{i+1}$. Let $\textbf{n}$ be a chosen unit normal vector of the face $f_i$. Using the right-hand rule, the normal $\textbf{n}$ provides the direction of the edges. We denote the directional edges by $\mathbf{e}_0$, $\mathbf{e}_1$, ..., $\mathbf{e}_{k-1}$ vectors. For the sake of simplicity, in the following explanation, we use a cyclic order of the edges, meaning $e_k$ and $e_{k+1}$ will also denote the edges $e_0$, and $e_1$, etc.

\begin{figure}[ht!]
    \centering
    \includegraphics[width = \columnwidth]{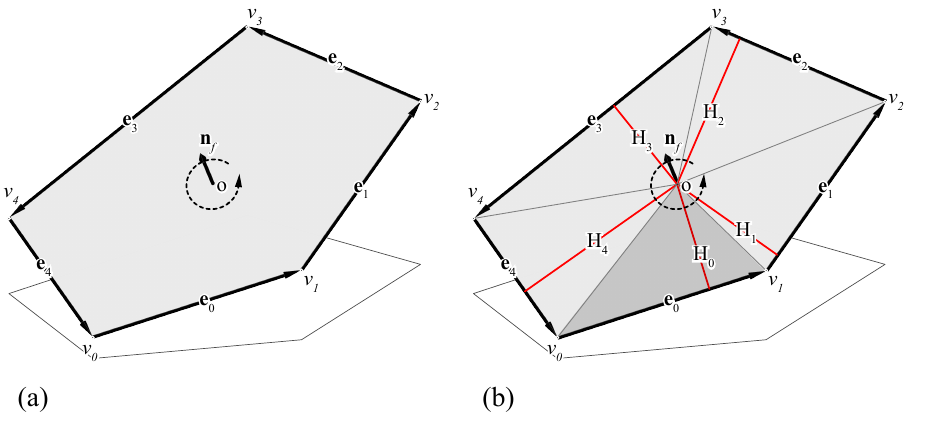}
    \caption{{(a) Vertices and edge vectors of the face $f$ with a normal direction $\textbf{n}_f$; and (b) dividing the face into triangles with a base $\textbf{e}_i$ and the height $H_i$ from the centroid of the vertices.}}
    \label{fig:area}
\end{figure}

We can compute the area of the face $f_i$ by :
\begin{itemize}
    \item dividing the face $f_i$ into $k$ triangles given by the $e_i$ and the geometric center of vertices, $O$; and
    \item compute the area of each triangle by computing its height $H_i$ which is the distance from $O$ to the edge $e_i$ (Figure \ref{fig:area}b).
\end{itemize}

\subsubsection*{Average height}

 The height $H_i$, the perpendicular distance of $O$ from the edge $e_i$, is the average of the signed projected distances $h_{i,j}$ of the vertices $v_0$, $v_1$, ..., $v_{k-1}$ from $e_i$ (Figure \ref{fig:area2}d).
 
 For instance, $h_{0,2}$, $h_{0, 3}$, and $h_{0, 4}$ are the signed distances of the vertices $v_2$, $v_3$ and $v_4$ from the edge $e_0$. The $h_{0,0}$ and $h_{0,1}$ are zero since the vertices $v_0$ and $v_1$ lay on edge $e_0$. Therefore, the area $A_i$ of the triangle $O$, $v_i$ and $v_{i+1}$ can be written as
\[A_i=\frac{1}{2}|\mathbf{e}_i|H_i=\frac{1}{2k}|\mathbf{e}_i|\sum_{j=0}^{k-1} h_{i,j},\]
and the total area of the face equals
\[A_f=\sum_{i=0}^{k-1} A_i=\frac{1}{2k}\sum_{0\leq i,j\leq k-1} |\mathbf{e}_i|h_{i,j}.\]

We are looking for a formula for the area $A_i$ based on the edge vectors. Thus,
let us compute the $h_{i,j}$ based on the edge lengths of the face $f_i$. Recall, that the vertices $v_i$ and $v_{i+1}$ are on the edge $e_i$, hence, $h_{i,i}$ and $h_{i,i+1}$ are zero (see Figure \ref{fig:area2}a, Figure \ref{fig:area2}d). The first vertex that can contribute non-trivially is $v_{i+2}$, and the height, $h_{i,i+2}$, can be computed by constructing the triangle of the vertices $v_i$, $v_{i+1}$ and $v_{i+2}$ (Figure \ref{fig:area2}a). We denote the signed area of this triangle by $A_{i,i+2}$ that can be computed using the two following methods: 
\begin{itemize}
 \item the area can be found by the height $h_{i,i+2}$ (Figure \ref{fig:area2} a):
    \begin{equation} \label{eq:area_tri_2}
    A_{i,i+2}=\frac{1}{2}|\mathbf{e}_i|h_{i,i+2}.
    \end{equation}

    \item also, the cross product of $\mathbf{e}_i$ and $\mathbf{e}_{i+1}$ provides the signed area:
    \begin{equation} \label{eq:area_tri_1}
        A_{i,i+2}\mathbf{n}=\frac{1}{2}(\mathbf{e}_i\times \mathbf{e}_{i+1})
    \end{equation}

Note that the sign of the area in Eq. \ref{eq:area_tri_2} is defined by the $h_{i,i+2}$ where it can only be negative in a concave or a self-intersecting polygon.

\begin{figure}[ht!]
    \centering
    \includegraphics[width = \columnwidth]{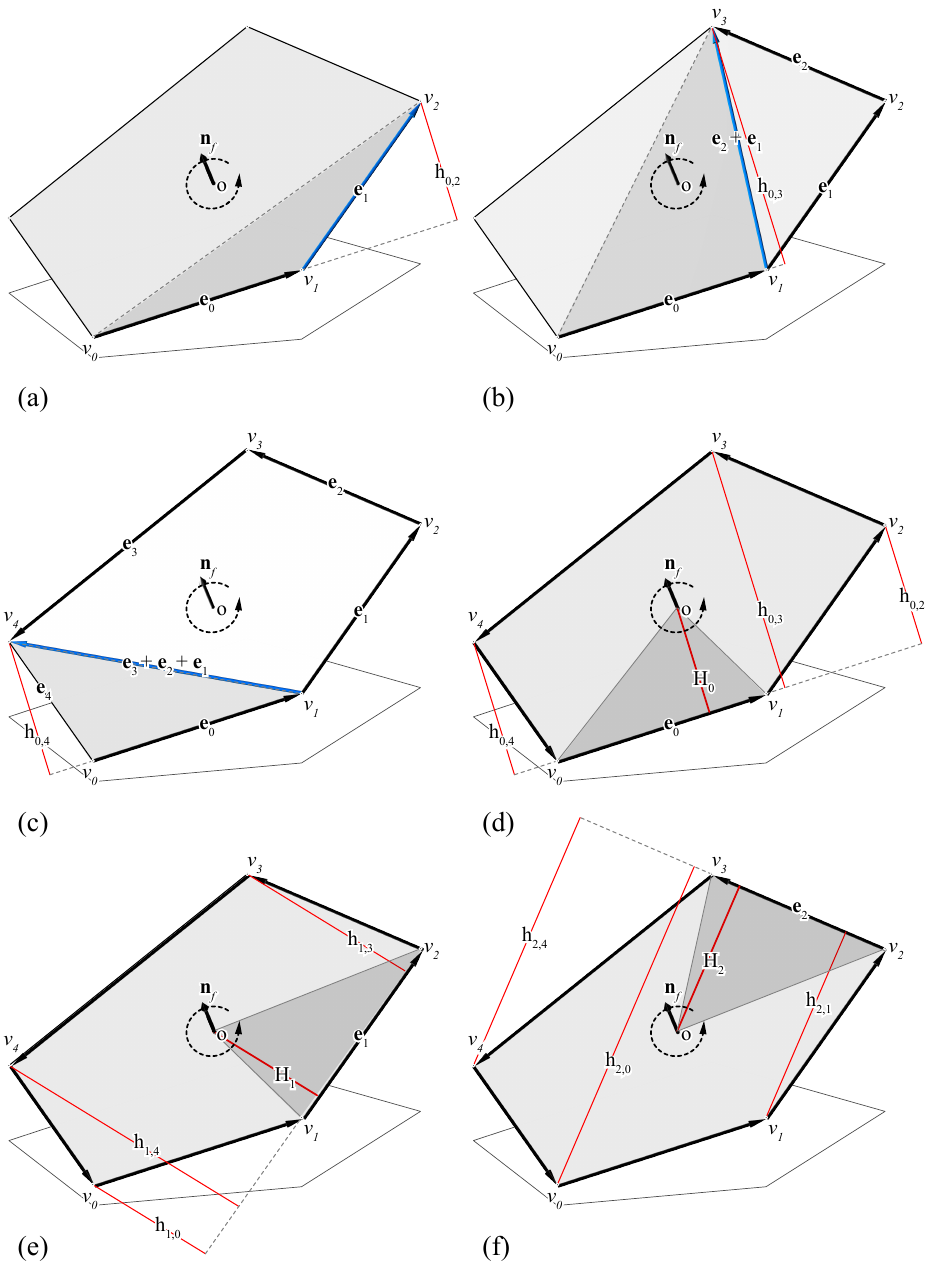}
    \caption{Finding the average height ${H}_i$ for each edge $\textbf{e}_i$ by constructing triangles starting from (a) $\textbf{e}_i$ and $\textbf{e}_{i+1}$, and (b)  $\textbf{e}_i$ and $\textbf{e}_{i+1} +
    \textbf{e}_{i+2}$ until we find all heights of the vertices in (c) and (d); and repeating the same process for other edges to find all ${H}_i$s in (e) and (f).}
    \label{fig:area2}
\end{figure}

\end{itemize}

From Eqs. \ref{eq:area_tri_2} and \ref{eq:area_tri_1}, we get the following
\begin{equation}\label{eq:vectordiv}\frac{1}{2}(\mathbf{e}_i\times \mathbf{e}_{i+1})=\frac{1}{2}|\mathbf{e}_i|h_{i,i+2}\mathbf{n}.\end{equation}

We have two vectors on the sides of the above equation. On one hand, it does not make sense to divide vectors by vectors. On the other hand, the vectors $\mathbf{e}_i$ and $\mathbf{e}_{i+1}$ are perpendicular to $\mathbf{n}$. Hence, the cross product of vectors $\mathbf{e}_i$ and $\mathbf{e}_{i+1}$ is either parallel or opposite to $\mathbf{n}$. Therefore, there exists a scalar $\mu_{i,i+1}$ so that
\[\mu_{i,i+1} \mathbf{n}=\mathbf{e}_i\times \mathbf{e}_{i+1}.\]
Thus, we can think of $\mu_{i,i+1}$ as 

 \begin{equation}\label{eq:mu}\mu_{i,i+1} = \frac{\mathbf{e}_i\times \mathbf{e}_{i+1}}{\textbf{n}}.\end{equation}


Going back to Eq. \ref{eq:vectordiv}, using the notations introduced above, we obtain a formula for $h_{i,i+2}$:
\begin{equation}\label{eq:tri3}h_{i,i+2}=\frac{\mathbf{e}_i\times \mathbf{e}_{i+1}}{|\mathbf{e}_i|\mathbf{n}}=\mu_{i,i+1}\frac{1}{ |\mathbf{e}_{i}|}.\end{equation}

The Eq. \ref{eq:tri3} is not of the form of a linear equation, because the scalar $\mu_{i,i+1}$ depends on the lengths of the edges $\textbf{e}_i$ and $\textbf{e}_{i+1}$. By dividing both sides of the Eq. \ref{eq:mu} by the edge lengths of $e_i$ and $e_{i+1}$, the scalar $\mu_{i,i+1}$ will become
\begin{equation}\label{eq:eta}
    \eta_{i,i+1}=\frac{1}{|\mathbf{e}_i|\cdot|\mathbf{e}_{i+1}|}\cdot \mu_{i,i+1}.
\end{equation}
The scalar $\eta_{i,i+1}$ will only depend on the directions of $\textbf{e}_i$ and $\textbf{e}_{i+1}$ and not on the edge lengths. Using this notation, we can write $h_{i,i+2}$ as
\begin{equation}\label{eq:hii+2}h_{i,i+2}=\eta_{i,i+1}|\mathbf{e}_{i+1}|.\end{equation}

In Eq. \ref{eq:hii+2}, the height is expressed as a product of a scalar $\eta_{i,i+1}$ that depends only on the direction of the edges, and the edge length of $\mathbf{e}_{i+1}$. Therefore, we obtained the height $h_{i,i+2}$ based on the linear function of the edge length.


In the next step, we will compute $h_{i,i+3}$. Consider the triangle with vertices $v_i$, $v_{i+1}$, $v_{i+3}$ (see Figure \ref{fig:area2}b). Note that the vector from $v_{i+1}$ to $v_{i+3}$ is the vector
\[\mathbf{e}_{i+1}+\mathbf{e}_{i+2}.\]

We compute the area $A_{i,i+3}$ of the triangle using two methods. The area can be computed by using the height $h_{i,i+3}$ (Figure \ref{fig:area2}b):
    \begin{equation*}
    A_{i,i+3}=\frac{1}{2}|\mathbf{e}_i|h_{i,i+3}.
    \end{equation*}

Also, the cross product of $\mathbf{e}_i$ and $\mathbf{e}_{i+1}+\mathbf{e}_{i+2}$ provides the signed area:
    \begin{equation*}
        A_{i,i+3}\mathbf{n}=\frac{1}{2}(\mathbf{e}_i\times (\mathbf{e}_{i+1}+\mathbf{e}_{i+2})).
    \end{equation*}

As a consequence, we obtain the following relationship.
\begin{equation*}
    \frac{1}{2}(\mathbf{e}_i\times (\mathbf{e}_{i+1}+\mathbf{e}_{i+2}))=\frac{1}{2}|\mathbf{e}_i|h_{i,i+3}\mathbf{n}
\end{equation*}
We combine this equation with Eq. \ref{eq:vectordiv} to obtain
\begin{equation}\label{eq:nii+2}
    \mathbf{e}_{i}\times \mathbf{e}_{i+2}+|\mathbf{e}_i|h_{i,i+2}\mathbf{n}=|\mathbf{e}_i|h_{i,i+3}\mathbf{n}
\end{equation}

Again, we would like to divide Eq. \ref{eq:nii+2} by the vector $\mathbf{n}$. Similarly, the vectors $\mathbf{e}_i$ and $\mathbf{e}_{i+2}$ are perpendicular to $\mathbf{n}$, hence $\mathbf{e}_i\times \mathbf{e}_{i+2}$ is either parallel or opposite to the direction of $\mathbf{n}$. We, again, introduce the scalar $\mu_{i,i+2}$ satisfying
\begin{equation*}
    \mu_{i,i+2}\mathbf{n}=\mathbf{e}_i\times \mathbf{e}_{i+2}.
\end{equation*}

With this notation, Eq. \ref{eq:nii+2} becomes
\begin{equation*}
    \mu_{i,i+2}+|\mathbf{e}_i|h_{i,i+2}=|\mathbf{e}_i|h_{i,i+3},
\end{equation*}
so we obtain a recursive formula
\begin{equation}\label{eq:hii+3}
    h_{i,i+3}=h_{i,i+2}+\frac{1}{|\mathbf{e}_i|}\mu_{i,i+2}.
\end{equation}
The scalar $\mu_{i,i+2}$ depends on the lengths of the edge vectors $\mathbf{e_i}$ and $\mathbf{e_{i+2}}$, however $\eta_{i,i+2}$ defined below does not:
\begin{equation*}
    \eta_{i,i+2}=\frac{1}{|\mathbf{e}_i|\cdot|\mathbf{e}_{i+2}|}\mu_{i,i+2}
\end{equation*}.
The scalar $\eta_{i,i+2}$ only depends on the directions of the edges. Using this notation and Eq. \ref{eq:hii+2}, Eq. \ref{eq:hii+3} becomes a linear expression for $ h_{i,i+3}$ 
\begin{equation}
    h_{i,i+3}=h_{i,i+2}+\eta_{i,i+2}|\mathbf{e}_{i+2}|=\eta_{i,i+1}|\mathbf{e}_{i+1}|+\eta_{i,i+2}|\mathbf{e}_{i+2}|
\end{equation}


Repeating this process for the rest of the edges (Figure \ref{fig:area2}c) results in a formula for $h_{i,i+l}$
\[h_{i,i+l}=\sum_{1\leq m\leq l-1}\eta_{i,i+m}|\mathbf{e}_{i+m}|.\]

Finally, we compute the average of these heights, $H_i$ by repeating the same process for all the edges of the face (Figure \ref{fig:area2}e,f):
\begin{align}
H_i&=\frac{1}{k}\sum_{j=2}^{k-1} h_{i,i+j}=\frac{1}{k}\sum_{j=2}^{k-1}\sum_{1\leq m\leq j-1}\eta_{i,i+m}|\mathbf{e}_{i+m}|=\\
&=\frac{1}{k}\sum_{j=1}^{k-2} (k-j-1)\eta_{i,i+j}|\mathbf{e}_{i+j}|
\end{align}

As a consequence, we obtain a quadratic formula for the area of the face $f$ in the edge lengths of the edges of $f$
\begin{equation}\label{eq:appendix}
    A_f=\frac{1}{2}\sum_{i=0}^{k-1} |\mathbf{e}_{i}|H_i=\frac{1}{2k}\sum_{i=0}^{k-1}\sum_{j=0}^{k-2} (k-j-1)\eta_{i,i+j}|\mathbf{e}_{i}||\mathbf{e}_{i+j}|.
\end{equation}

\subsubsection*{Quadratic form}
The next step is to turn the quadratic equation \ref{eq:appendix} into a quadratic form with a matrix. Note, that we can compute the coefficients
\[(k-j-1)\eta_{i,i+j}\]
in Eq. \ref{eq:appendix} without knowing the edge lengths. As a result, we can formulate the right-hand side of the Eq. \ref{eq:appendix} in a quadratic form given by a matrix $\textbf{M}_f$, whose entries are given by the coefficients:
\begin{equation}\label{eq:matrixcoeff}
\textbf{M}_{f,ij}=\begin{cases} (k-j+i-1)\eta_{i,j} & \mbox{if }j > i\\
 (i-j-1)\eta_{i,j} & \mbox{if }j < i\\
 0 & \mbox{if }i=j
 \end{cases}.
 \end{equation}

Thus, we can rewrite Eq. \ref{eq:appendix} as a quadratic form
 \begin{equation}\label{eq:quadrform}
 A_{f}=\frac{1}{2k}\textbf{q}^T\textbf{M}_{f}\textbf{q}
 \end{equation}
 where $\mathbf{q}$ is the column vector of the edge lengths
 \[\mathbf{q}=(|\mathbf{e}_1|,|\mathbf{e}_2|,...,|\mathbf{e}_k|)^T.\]

\subsubsection*{Symmetric matrix}
Usually, the matrix  $\mathbf{M}_f$ is not a symmetric matrix. However, the computations may become simpler if the matrix is symmetric. Indeed, the matrix $\mathbf{M}_f$ can be turned into a symmetric matrix. Since 
\begin{equation*}
    \left(\mathbf{q}^T\mathbf{M}_f\mathbf{q}\right)^T=\mathbf{q}^T\mathbf{M}^T_f\mathbf{q}
\end{equation*}
the quadratic form given as
\[\frac{1}{2k}\mathbf{q}^T\mathbf{M}^T_f\mathbf{q}\] 
also computes the area of the face. 

As a consequence, 
\begin{equation*}
    A_f=\frac{1}{2k}\mathbf{q}^T\frac{\mathbf{M}_f+\mathbf{M}^T_f}{2}\mathbf{q},
\end{equation*}
and in this case, the corresponding matrix
\begin{equation}\label{eq:symmat}
    \frac{\mathbf{M}_f+\mathbf{M}^T_f}{2}
\end{equation}
is symmetric. 

For the sake of computational simplicity, from now on, we assume that the matrix $\mathbf{M}_f$ appearing in Eq. \ref{eq:appendix} is always symmetric.

%% file: method_03.tex
\subsection{Computing the face geometry for a target area}\label{sec:face}
In this section, we develop a method to reconstruct a given face $f_i$ by constraining particular user-defined edge lengths and the target area for the face. To give a general overview of our approach, consider the face $f_i$ of Figure \ref{fig:area}a. Initially, without any constraint, we have five unknowns which are the edge lengths of the five edges $e_{0-4}$. There are three equilibrium equations based on Eq.\ref{eq:eqface}, in $-x$, $-y$, $-z$ around the face $f_i$, and one of them is redundant \cite{akbarzadeh2018developing}. As a consequence, the dimension of the possible solutions, i.e. the possible faces, satisfying the equilibrium equations is 
 \[e-2 = 3.\]

 Instead of solving the quadratic area equation \ref{eq:quadrform} for three unknowns, we constrain two of them and solve the quadratic equation for only one unknown. Using this technique, we can significantly reduce the complexity of finding a solution for this quadratic equation. This provides additional design possibilities for the user, as we either allow the user to define up to two edge lengths out of three or we use the existing edge lengths for two unknown edges and compute the area based on the last unknown edge. 
 



In general, our goal is to simplify solving the quadratic equation of the face by solving it for only one edge length. 

\subsubsection*{Computing $GDoF_f$ using RREF}

The dimension of the possible solutions for the geometry of the face is called the Geometric Degrees of Freedom ($GDoF_f$). In fact, $GDoF_f$ describes the dimension of the family of polygons with edges parallel to the edges of the initial face which is always equal to:
\[e-2.\]

The GDoF is also equal to the number of independent edges in each face. In fact, the lengths of the independent edges can define the lengths of the rest of the edges and the geometry of the face \cite{akbarzadeh2020geometric}. The independent edges can be found using the Reduced Row Echelon form method (RREF). 

Specifying the edge lengths for the $e-2$ independent edges yields a unique solution for the geometry of the face. Obviously, this solution does not solve the area equation. However, we can specify $e-3$ independent edges. In that case, we will have infinitely many solutions given by the edge length of the last independent edge that, in fact, provides the possibility to solve the area equation. 

This method provides a solution to recompute the geometry of the face with a given target area. However, the objective of the research is to construct the geometry of a face with a given target area and user-defined edge lengths.

\subsubsection*{Constrained Geometric Degrees of Freedom ($CGDoF_f$)}

The user-defined edge lengths provide linear equations for the edge lengths that are in general non-homogeneous. As a consequence, the dimension of the solution space for possible geometries of the face may decrease significantly. 

The user may over-determine the system, for instance, by assigning too many edge lengths. To avoid this problem, we compute the dimension of the constrained solution space, called the Constrained Geometric Degrees of Freedom (CGDoF), using RREF. 

The result of this method classifies the edges into the following classes: 
\begin{itemize}
    \item the fixed edges, $e_{fix}$: the edges chosen by the user with predefined edge lengths (these edges are always dependent edges of the equation system);
    \item the non-fixed dependent edges, $e_{nfd}$: the dependent edges which are not predefined by the user, and
    \item the independent edges, $e_{ind}$.
\end{itemize}
To solve the quadratic equation for the area, we reduce the number of independent edges $e_{ind}$ to one, by assigning the existing edge length for all independent edges except one. The last remaining independent edge is called the critical independent edge, $e_{ci}$, the length of which we find using the quadratic equation. This method will be described in detail in the next sections.

\subsubsection*{Defining the Constrained equations for a face}\label{sec:defce}
In Section \ref{sec:area}, we expressed the area of a face polygon based on a quadratic form of the edge lengths. Now, we develop linear, non-homogeneous constrained equations describing the geometry of the face with preassigned lengths for certain edges of the face.

We can write the edge $\textbf{e}_i$ with a predefined length $q_i$ as a constraint vector equation in the following way:
\begin{equation}\label{eq:eqlf}
   \mathbf{l}_i^T\mathbf{q}=q_i
\end{equation}
where $\mathbf{l}_i$ is the $[e\times 1]$ column vector whose entries are all zero $(0)$ except at the index of $e_i$ where it is one $(1)$.

Similarly, multiple constraints, i.e other fixed edge lengths, can be written as a matrix equation 
\begin{equation}\label{eq:eq_L_f}
    \mathbf{L}_f\mathbf{q}=\mathbf{l}_f
\end{equation}
where the rows of $\mathbf{L}_f$ are the row vectors $\mathbf{l}_i^T$ and $\mathbf{l}_f$ is the vector whose entries are the $q_i$, the predefined edge lengths.


Together with the equilibrium equations of \ref{eq:eqface}, we obtain all the linear equations describing the linear constraints which results in the constraint equation system
\begin{equation}\label{eq:cons}
    \mathbf{B}_{f}\mathbf{q}=\mathbf{b}_f
\end{equation}
where the matrix $\mathbf{B}_f$ is obtained by stacking the matrices $\mathbf{E}_f$ and $\mathbf{L}_f$
\begin{equation*}
    \mathbf{B}_f=\left(\begin{array}{c}
         \mathbf{E}_f  \\
         \mathbf{L}_f
    \end{array}
    \right),
\end{equation*}

and the vector $\mathbf{b}_f$ is obtained as stacking the zero vector and the vector $\mathbf{l}_f$ together
\begin{equation*}
    \mathbf{b}_f=\left(\begin{array}{c}
         \mathbf{0}\\
         \mathbf{l}_f
    \end{array}
    \right)
\end{equation*}
We call the matrix $\mathbf{B}_f$ the constraint matrix and the vector $\mathbf{b}_f$ the constraint vector.

\subsubsection*{Analyzing the constraint equation system (RREF)}

The constraint equation system, Eq. \ref{eq:cons}, is, in general, a non-homogeneous, linear equation system. The solution space of this equation system is often not a linear subspace but rather the empty set or an affine subspace of the possible solution space $\mathbb{R}^e$. Here, $e$ denotes the number of edges. The dimension of this affine subspace is the Constrained Geometric Degrees of Freedom of the face ($CGDoF_{f}$). The $CGDoF_{f}$ is the geometric degrees of the face after applying the edge constraints by the user. 

It is also possible to have no solutions for Eq. \ref{eq:cons}. In this case, we say that the $CGDoF_f$ is $-\infty$. If there exists a solution to Eq. \ref{eq:cons}, then the $CGDoF_f$ is a non-negative integer, which is the dimension of the affine subspace formed by the solutions. When the $CGDoF_f$ is zero, the constraint equations have a unique solution. If the $CGDoF_f$ is positive, then there are many significantly different solutions to the constraint equations.


In order to compute the $CGDoF_f$, we use the reduced row echelon form ($\textbf{RREF}_f$) of the matrix obtained from the constraint matrix, $\mathbf{B}_f$, and the constraint vector $\mathbf{b}_f$: 
\begin{equation*}
    \left(\begin{array}{c|c}
        \mathbf{B}_f &  \mathbf{b}_f
    \end{array}
    \right).
\end{equation*} 
The $CGDoF_f$ can be easily computed from this reduced- row-echelon form, but the following two possibilities might occur:
\begin{itemize}
    \item If there exists a row of $\textbf{RREF}_f$, so that the last entry is one, but all other entries are zero: 
    \[\left(\begin{array}{ccc|c}
        0 & ...& 0 & 1
    \end{array}
    \right)\]
    then, the constraint equation, Eq. \ref{eq:cons} have no common solution. In this case, the $CGDoF_f$ is $-\infty$, and the user needs to modify the constraints and/or release some of the constrained edges from his/her input.
    
    \item Otherwise, we have at least one solution. In this case, the $CGDoF_f$ equals $e-r$ where $e$ is the number of edges of the face and the number of columns of the constraint matrix $\textbf{B}_f$, and $r$ is the rank of the $\textbf{RREF}_f$. This rank equals the number of pivots that also equals the rank of $\mathbf{B}_f$.
\end{itemize}

\subsubsection*{Solving the area equation}
The main idea of manipulating the edge lengths of the face to obtain a required area is to solve the area equation Eq. \ref{eq:quadrform} by reducing the number of unknown edges into a single unknown edge length. 

We reduce the unknowns by finding all the independent edges of the constraint equation system \ref{eq:cons} and assigning either the current values or a user-defined values to them. 

From now on, we assume that there exists a solution to Eq. \ref{eq:cons}. The columns in $\textbf{RREF}_f$ corresponding to the pivots are called the pivotal columns. The non-pivotal columns correspond to the so-called independent edges whose lengths can be manipulated freely. Once the values for the independent edges are set (possibly by the user), there is a unique solution to Eq. \ref{eq:cons}.

The edges corresponding to the pivotal columns are the edges depending on the independent ones, these edge lengths will be updated so that the Eq. \ref{eq:cons} is satisfied. 

Our method for solving the area Eq. \ref{eq:quadrform} is to set as many values of the lengths of the independent edges as possible. In this case, it is one less than the number of independent edges: $e-r-1$. The length of the last independent edge, $q_{ci}$, is the length of the \textit{critical independent} edge, $e_{ci}$. This length will be treated as a variable for which we will solve the equation.
 
To solve Eq. \ref{eq:quadrform}, we organize the edges according to the form of $\textbf{RREF}_f$ into vectors:
\begin{itemize}
    \item the vector corresponding to the critical edge is defined as an $[e\times 1]$ column vector $\mathbf{q}_{ci}$:
    \begin{equation*}
        \mathbf{q}_{ci, i}=\begin{cases} q_{ci} & \mbox{if i is the index of the $e_{ci}$ }\\
        0 & \mbox{otherwise}
        \end{cases}
    \end{equation*}
    Here $q_{ci}$, the edge length of the $e_{ci}$ edge, is the unknown and we will solve the area equation for $q_{ci}$;
    \item the vector of the edge lengths of the \textit{non-critical independent} edges, $e_{nci}$, is defined as $\mathbf{q}_{nci}$ which is an $[e\times 1]$ column vector:
    \begin{equation*}
        \mathbf{q}_{nci}=\begin{cases} q_i & \mbox{if i is the index of an $e_{nci}$ edge}\\
        0 & \mbox{otherwise}
        \end{cases}
    \end{equation*}
    where $q_i$ are the current edge lengths of the $e_{nci}$ edges;

    \item Similarly, the vector of the edge lengths of the fixed/predefined edges, $e_{fix}$ is defined by $\mathbf{q}_{fix}$ as 
    \begin{equation*}
        \mathbf{q}_{fix}=\begin{cases} q_i & \mbox{if i is the index of a user-selected edge}\\
        0 & \mbox{otherwise}
        \end{cases}
    \end{equation*}
    The indices of fixed edges are indices of some of the pivotal columns. These $q_i$ are fixed in the beginning of the problem and won't be updated;
    
    \item Finally, the vector of edge lengths of the \textit{non-fixed dependent} edges, $e_{nfd}$ is defined by vector $\mathbf{q}_{nfd}$ as
    \begin{equation*}
         \mathbf{q}_{nfd}=\begin{cases} q_i & \mbox{if i is the index of an $e_{ndf}$ }\\
        0 & \mbox{otherwise}
        \end{cases}
    \end{equation*}
    The edge lengths $q_i$ of the $e_{nfd}$ edges can be computed from the lengths of the edges corresponding to $e_{ci}$ , $e_{nci}$, and $e_{fix}$. The edge lengths will be updated in order to satisfy the area equation.
\end{itemize}

 \begin{figure}
     \centering
     \includegraphics[width = 1\columnwidth]{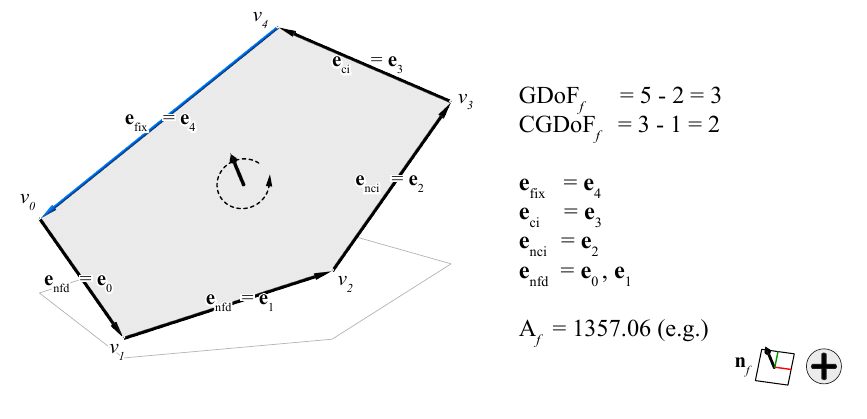}
     \caption{A sample face showing the edge vectors, its normal direction and the choices of $e_{fix}$, $e_{cr}$, $e_{nci}$, and $e_{nfd}$.}
     \label{fig:fig_12_01}
 \end{figure}

After setting up the notations, we begin to solve the area Eq. \ref{eq:quadrform}.

Since any edge is either $e_{ci}$, $e_{nci}$, $e_{fix}$ or $e_{nfd}$, we have an equality
\begin{equation}\label{eq:lotq}
\mathbf{q}=\mathbf{q}_{ci}+\mathbf{q}_{nci}+\mathbf{q}_{fix}+\mathbf{q}_{nfd}.\end{equation}
Moreover, the lengths of the $e_{nfd}$ depend linearly on the lengths of the $e_{ci}$, $e_{nci}$ and $e_{fix}$, hence there exist an $[e\times e]$ square matrix, $\mathbf{D}$ and vectors $\mathbf{d}$, $\mathbf{g}$ so that:
\begin{equation}\label{eq:dep}
\mathbf{q}_{nfd}=\mathbf{D}\mathbf{q}_{nci}+\mathbf{d}q_{ci}+\mathbf{g}
\end{equation}
The matrix $\mathbf{D}$ and the vectors $\mathbf{d}$ and $\mathbf{g}$ can be computed from the RREF form easily as follows. 

The matrix $\mathbf{D}$ is a matrix whose entries are mostly 0 except at the entries corresponding to the columns of $e_{nci}$ and to the rows of the $e_{nfd}$, where the value is the opposite of the corresponding value in the $\textbf{RREF}_f$ matrix of $\left(\begin{array}{c|c}
        \mathbf{B}_f &  \mathbf{b}_f
    \end{array}
    \right).$

 The vector $\mathbf{d}$ is a vector whose entries are mostly zero (0) except at the entries corresponding to the column of the $e_{ci}$ edge and to rows of the dependent edges, where the value is the opposite of the corresponding value in the $\textbf{RREF}_f$ matrix of $\left(\begin{array}{c|c}
        \mathbf{B}_f &  \mathbf{b}_f
    \end{array}
    \right).$

The vector $\mathbf{g}$ is the contribution coming from the fixed edges. This is a vector whose entries are mostly zero (0), except for entries corresponding to the indices of the $e_{nfd}$ edges, when the entry is the last entry of the corresponding row of the $\textbf{RREF}_f$ matrix $\left(\begin{array}{c|c}
        \mathbf{B}_f &  \mathbf{b}_f
    \end{array}
    \right).$

We simplify Eq \ref{eq:lotq} slightly. Consider the $[e\times e]$ square matrix $\mathbf{Id}_{nci}$, which is the identity restricted to the $e_{nci}$ edges, and 0 elsewhere. Since, 
\begin{equation*}
    \mathbf{Id}_{nci}\mathbf{q}_{nci}=\mathbf{q}_{nci}
\end{equation*}
we have that
\begin{equation*}\label{eq:q_dep}
\mathbf{q}_{nfd}+\mathbf{q}_{nci}=\mathbf{D}'\mathbf{q}_{nci}+\mathbf{d}q_{ci}+\mathbf{g}
\end{equation*}
where $\mathbf{D'}$ is the matrix $\mathbf{D}+\mathbf{Id}_{nci}$.

Thus, by Eq. \ref{eq:lotq}, we have
\begin{equation*}
    \mathbf{q}=\textbf{q}_{ci}+\mathbf{D}'\mathbf{q}_{nci}+\mathbf{q}_{fix}+\mathbf{d}q_{ci}+\mathbf{g}.
\end{equation*}
Let us denote $\mathbf{d'}$ by the vector obtained by adding a 1 to the vector $\mathbf{d}$ at the index of the critical edge, i.e $\mathbf{d'}q_{ci}=\mathbf{q}_{ci}+\mathbf{d}q_{ci}$. Then, we have
\begin{equation}\label{eq:qform}
    \mathbf{q}=\mathbf{D'}\mathbf{q}_{nci}+\mathbf{q}_{fix}+\mathbf{d'}q_{ci}+\mathbf{g}
\end{equation}

Now, we can solve the area Eq. \ref{eq:quadrform} by plugging in the right-hand side of Eq. \ref{eq:qform} into $\mathbf{q}$: the quantity
\begin{equation}
\resizebox{0.85\hsize}{!}{
    $\frac{1}{2k}\left(\mathbf{D'}\mathbf{q}_{nci}+\mathbf{q}_{fix}+\mathbf{d'}q_{ci}+\mathbf{g}\right)^T \textbf{M}_f \left(\mathbf{D'}\mathbf{q}_{nci}+\mathbf{q}_{fix}+\mathbf{d'}q_{ci}+\mathbf{g}\right)$}
\end{equation}
computes the area of the face, $A_f$. Rearranging the terms, we obtain a quadratic equation for $q_{ci}$:
\begin{equation}\label{eq:quadr}
    aq_{ci}^2+bq_{ci}+c=0
\end{equation}
where
\begin{equation*}
    a=\mathbf{d'}^T\mathbf{M_f}\mathbf{d'}
\end{equation*}
and
\begin{equation*}
    b=2\mathbf{d'}^T\mathbf{M_f}\left(\mathbf{D'}\mathbf{q}_{nci}+\mathbf{q}_{fix}+\mathbf{g}\right)
\end{equation*}
and
\begin{equation*}
    c=\left(\mathbf{D'}\mathbf{q}_{nci}+\mathbf{q}_{fix}+\mathbf{g}\right)^T\mathbf{M_f}\left(\mathbf{D'}\mathbf{q}_{nci}+\mathbf{q}_{fix}+\mathbf{g}\right)-2kA_f
\end{equation*}

We can solve this quadratic equation (using the quadratic formula) to obtain possibly two solutions for $q_{ci}$:
\begin{equation*}
    q_{ci}=\frac{-b\pm\sqrt{b^2-4ac}}{2a}
\end{equation*}

\subsubsection*{Number of solutions}\label{sec:number}
\begin{figure}
    \centering
    \includegraphics[width = \columnwidth]{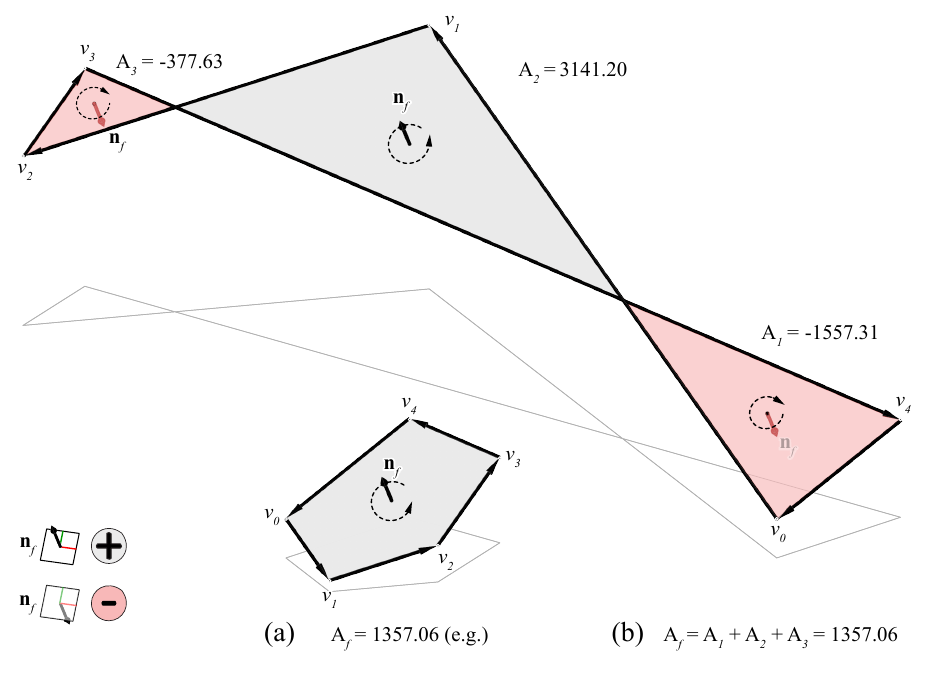}
    \caption{There are at least two significantly polygon different geometries that represent the same area of a polygon.}
    \label{fig:two}
\end{figure}

Depending on the target area, $A_f$, we might have different number of solutions for Eq. \ref{eq:quadr}. It is possible to have no solution, a unique solution, or two significantly different solutions (see Figure \ref{fig:two}).

Depending on the signature of $A$, a large positive or a small negative prescribed area ensures that we have multiple (two) solutions.

\subsubsection{Updating the edges of the face}\label{sec:updatefi}
\begin{figure}
    \centering
    \includegraphics[width = \textwidth]{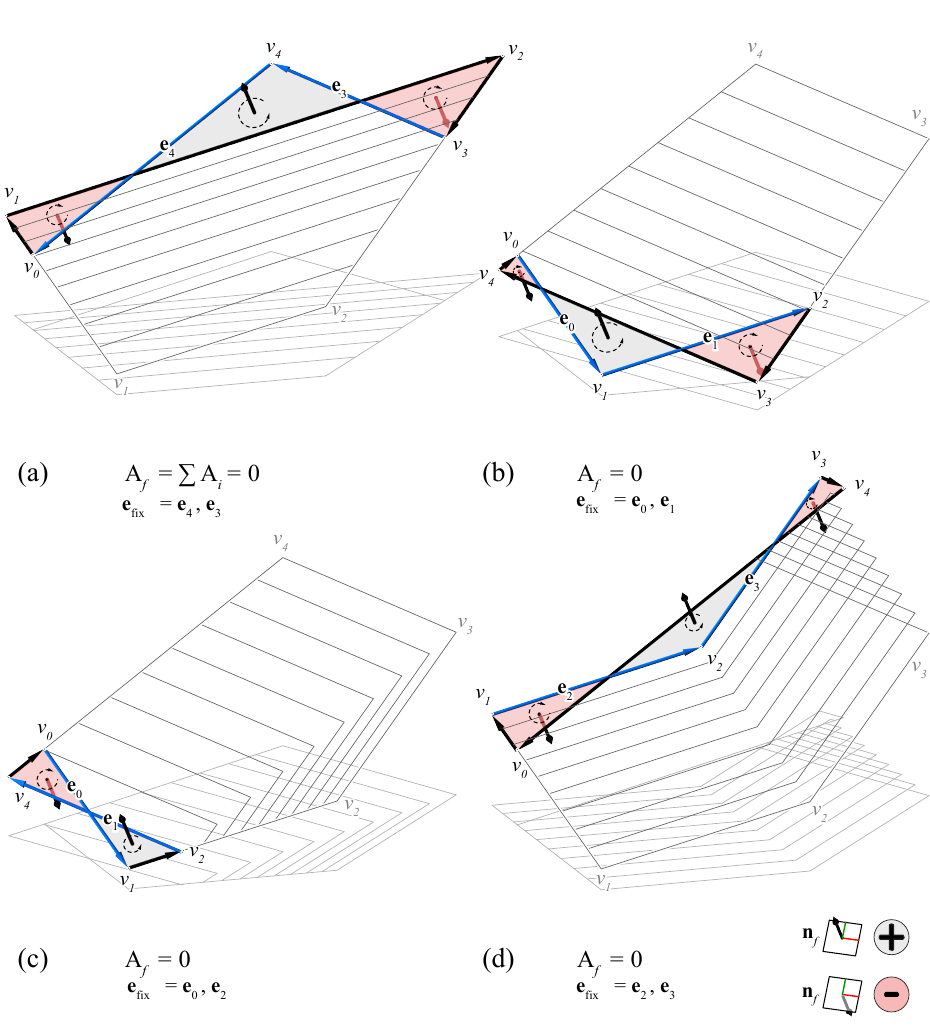}
    \caption{(a) to (d) various zero-area computation for a starting face with the area $A_f$ and different chosen fixed edges.}
    \label{fig:face_comp1}
\end{figure}

Once we computed the length $q_{ci}$ of the $e_{ci}$ edge, we can update the lengths of the dependent edges using Eq. \ref{eq:dep}. Now, all the lengths of the edges are computed. The face corresponding to the edge lengths has the required area and the edges of the face satisfy the constraint equation system, Eq. \ref{eq:cons} while only the lengths of the $e_{nfd}$ edges and the length of the $e_{ci}$ were manipulated (Figure \ref{fig:face_comp1}). 

The above discussion is summarized in Algorithm \ref{alg:AF} in Section \ref{sec:impf}.

\begin{figure}
    \centering
    \includegraphics[width = \columnwidth]{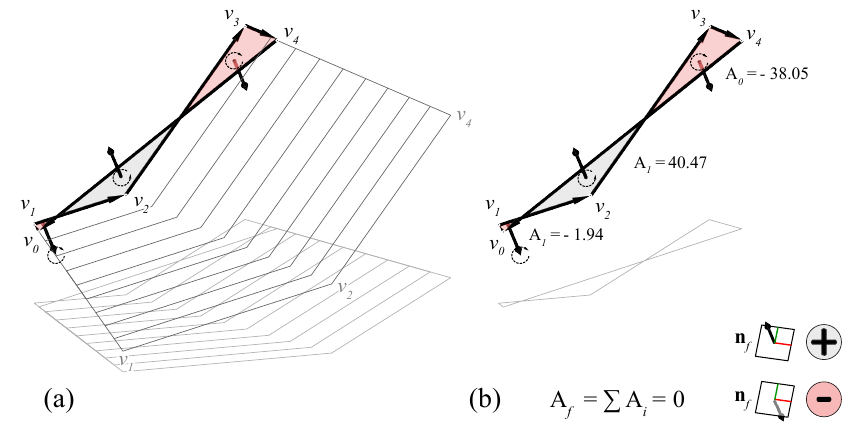}
    \caption{(a) Multiple steps to compute the new face geometry with the preassigned area (for visualization purposes only); and (b) the computed face geometry with zero area.}
    \label{fig:face_comp0}
\end{figure}

\subsubsection{Example}

Consider the pentagon of Figure \ref{fig:area}. The matrix $\mathbf{M}_f$ for this example is the following:
  
\[\left(\begin{array}{ccccc}0 &          -2.872 &          -1.85 & 0.506 & 0 \\
0 &  0 &         -1.893 &          -1.358 & 0.38\\
 0.925  & 0  & 0 &          -2.97 &          -0.577\\
  -1.012  & 0.679  & 0  & 0 &          -2.722        \\
  -2.986 &          -0.761 & 0.288 & 0 & 0        \end{array}\right)\]
    
Here, the order of the edges is given as $e_{(0,1)}$, $e_{(1,2)}$, $e_{(2,3)}$, $e_{(3,4)}$ and $e_{(4,0)}$. The user selects $e_{(4,0)}$ as the fixed edge and assigned area of the face to be zero. 

The matrix $\mathbf{B}_f$ is given as
\[\left(\begin{array}{ccccc}
0.289 & 1 & 0.776 & -0.734 & -0.925\\
-0.957 & 0 & 0.631 & 0.679 & -0.38\\
0 & 0 & 0 & 0 & 1
\end{array}\right)\]
and the vector $\mathbf{b}_f$ is
\[\left(\begin{array}{c}
0\\
0\\
41.78
\end{array}\right).\]
Here the first two rows describe the equilibrium equations, Eq. \ref{eq:eqface}, --for the $x-$ and $y-$coordinates. The third equation is the constraint equation, Eq. \ref{eq:eq_L_f}, for the fixed edge whose length is $41.78$. In our case, the fixed edge corresponds to the last entry.

The $\textbf{RREF}_f$ matrix is
\begin{equation}\label{eq:exarref}
    \left(\begin{array}{ccccc|c}
1 & 0 & -0.659 & -0.709 & 0 & -16.602\\
0 & 1 & 0.966 & -0.529 & 0 & 43.441\\
0 & 0 & 0 &0 &1 & 41.78 
\end{array}\right).
\end{equation}
As a consequence, the $CDGoF_f$ can be computed as $e-r=5-3=2$.


Moreover, we can identify the $e_{ci}$, $e_{nci}$, $e_{fix}$ and $e_{nfd}$ edges from the RREF matrix, \ref{eq:exarref} as follows.

The fixed edge $e_{(4,0)}$ corresponds to the fifth entry. Since, the $CDGoF_f$ equals to $2$, we have two more ($5-2-1=2$) dependent edges. These are the $e_{nfd}$ edges, $e_{(0,1)}$ and $e_{(1,2)}$, given by the other pivotal columns. The $e_{ci}$ edge was chosen to be the edge corresponding the fourth entry, $e_{(3,4)}$. Finally, the $e_{nci}$ edge is the remaining edge, $e_{(2,3)}$ for which we solve the quadratic area equation (see Figure \ref{fig:fig_12_01}). 


Now, we compute the coefficients of Eq. \ref{eq:quadr} to solve for the edge length, $q_{ci}$ of the $e_{ci}$ edge. Here, the target area, $A_f$ is zero.
\[a=-1.796,
b=-390.646,
c=-1898.751\]
We obtain two solution for Eq. \ref{eq:quadr}
\[q_{ci}=-4.974\mbox{ and } 212.535.\]
As a consequence, we get two significantly different solutions for the geometry of this face. The updated (self-intersecting) face corresponding to $q_{ci}=-4.974$ is shown in Figure \ref{fig:face_comp0}.

%% file: method_04.tex
\subsection{Computing the polyhedral geometry for target areas}\label{sec:poly}
In this process, a user would select multiple internal/external faces and edges of a polyhedral system and would assign target areas for each face and edge lengths for each edge to compute the new geometry of the polyhedron. Computing the geometry of a system of polyhedral cells with pre-assigned areas and fixed edge lengths in one step is a complex task. We propose a multi-step, inductive process to tackle this problem.

\begin{figure*}
    \centering
    \includegraphics[height = .9\textheight]{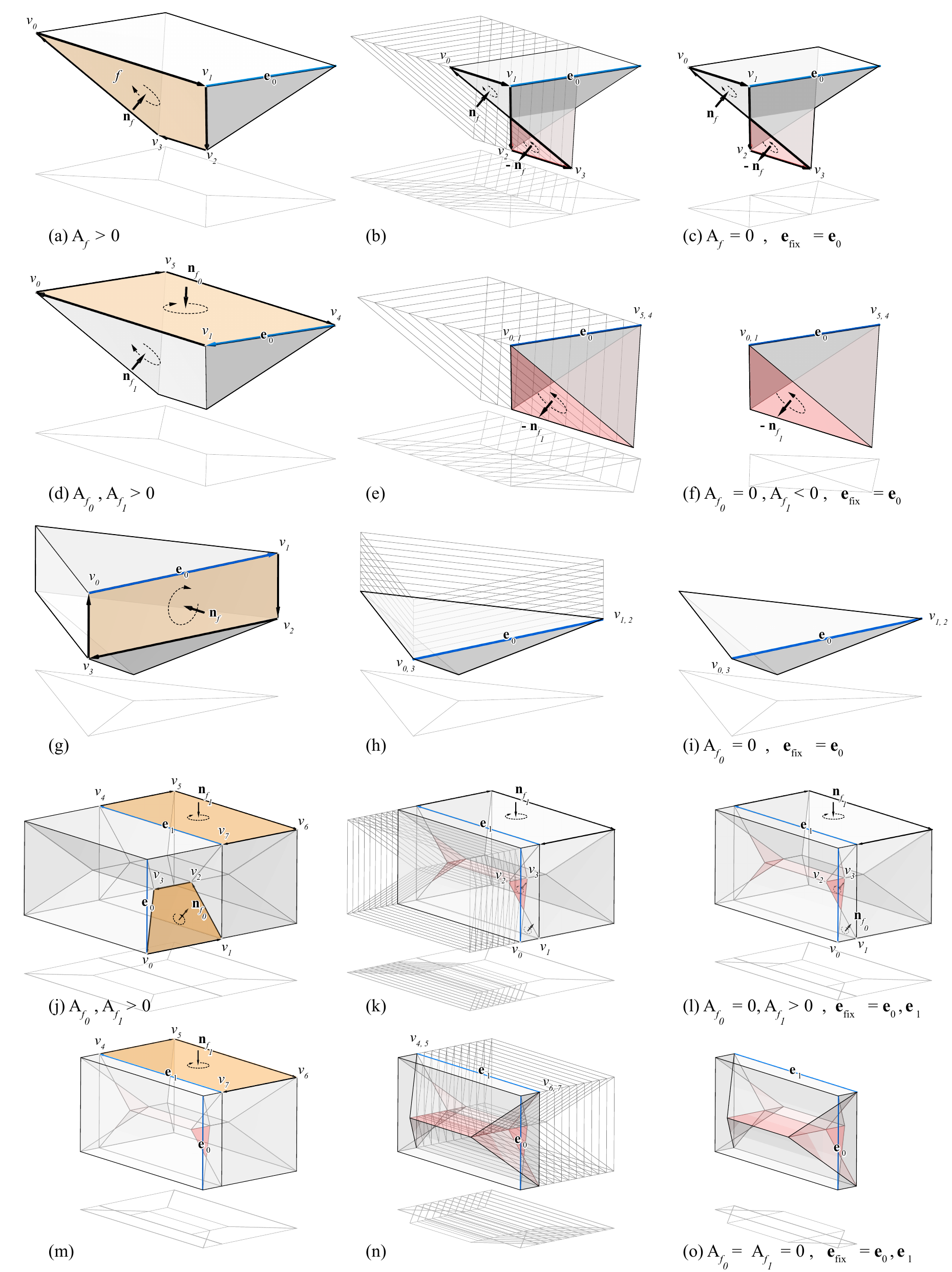}
    \caption{(a), (d), (g), (j), (m) Multiple polyhedral geometries with selected faces (orange) and user-assigned fixed edges; (b), (e), (h), (k), (n) the face area computation and visualization in multiple steps (for visualization purposes only); and (c), (f), (i), (l), and (o) the resulting polyhedral geometries with zero areas for the selected faces.}
    \label{fig:face_comp}
\end{figure*}





\subsubsection{Prescribed area for one face}\label{sec:prearea}
In each step, we compute the geometry of a single face $f_i$ with the assigned area as described in Section \ref{sec:face}. Then we update the polyhedron with the new edge lengths using Moore-Penrose Inverse (MPI) Method and move to the next face and repeat this process until there is no face left to change.

First, we identify the fixed edges from the list $e_{fix}$ which lie on the face $f_i$ with prescribed edge lengths, and use RREF method to solve the quadratic area Equation \ref{eq:quadrform} described in Section \ref{sec:face}.

In the next step, to preserve the new geometry of the face $f_i$, we consider all the newly generated edges of the face as fixed edges for the entire polyhedral system. I.e., we update the list of fixed edges $e_{fix}$ for the entire polyhedron with the newly computed edge lengths of  $f_i$.

 \subsubsection{Non-homogeneous equation system for a polyhedron}
 
 Similarly to Section \ref{sec:defce}, Eq. \ref{eq:cons}, the linear constraint equations for a polyhedral system can be described by two different kinds of linear equations. First, we have the equilibrium equation system, Eq. \ref{eq:eqpoly} describing the topology of the polyhedral system. Second, we have the linear constraints coming from the prescribed edge lengths.
 
 As a result, we obtain an equation system that describes the equilibrium of the polyhedron with the prescribed edge lengths of the fixed edges:
\begin{equation}\label{eq:cons2}
    \mathbf{B}_p\mathbf{q}=\mathbf{b}_p
\end{equation}
where the constraint matrix 
\begin{equation*}
    \mathbf{B}_p=\left(\begin{array}{c}
         \mathbf{E}_p  \\
         \mathbf{L}_p
    \end{array}
    \right)
\end{equation*}
is built from the equilibrium matrix of the polyhedron $\textbf{E}_p$ (see Eq. \ref{eq:eqpoly}) and the constraint equations $\mathbf{L}_p\mathbf{q}=\mathbf{l}_p$ coming from the fixed edges (see Eq. \ref{eq:eq_L_f}). Similarly, the vector \begin{equation*}
    \mathbf{b}_p=\left(\begin{array}{c}
         \mathbf{0}\\
         \mathbf{l}_p
    \end{array}
    \right)
\end{equation*}
is obtained from the edge vector of the fixed edges $\mathbf{l}_p$.

\subsubsection{Solving non-homogeneous equation systems using MPI}\label{sec:mpi}
Now, we propose to solve Eq. \ref{eq:cons2} using the Moore-Penrose Inverse (MPI) method. The MPI method is a technique to solve a general non-homogeneous equation system of the matrix form
\begin{equation}\label{eq:mpi}
    \mathbf{B}_p\textbf{q}=\mathbf{b}_p
\end{equation}
where the matrix $\mathbf{B}_p$ and the vector $\mathbf{b}_p$ are given.

We represent MPI of the matrix $\textbf{B}_p$ by $\mathbf{B}_p^{+}$ that satisfies the following matrix equations

\begin{equation*}
\textbf{B}_p\textbf{B}_p^{+}\textbf{B}_p=\textbf{B}_p \quad\mbox{and}\quad \textbf{B}_p^{+}\textbf{B}_p\textbf{B}_p^{+}=\textbf{B}_p^{+}.
\end{equation*}

Assume that the vector $\textbf{b}_p$ is of the form $\textbf{B}_p\textbf{q}$ for some vector $\textbf{q}$. Multiplying the first equality by $\textbf{q}$, we have
\[\textbf{B}_p\textbf{B}_p^+\textbf{B}_p\textbf{q}=\textbf{B}_p\textbf{q}.\]
As a consequence, we obtain
\begin{equation}\label{eq:forb}
    \textbf{B}_p\textbf{B}_p^+ \textbf{b}_p=\textbf{b}_p.
\end{equation}
Therefore, if a solution to Eq. \ref{eq:mpi} exists, then the Eq. \ref{eq:forb} has to be satisfied. Similarly, if Eq. \ref{eq:forb} is satisfied, then the vector $\mathbf{B}_p^+ \mathbf{b}_p$ is a solution to Eq. \ref{eq:mpi}
This provides an effective tool to check whether Eq. \ref{eq:mpi} has a solution or not.

From now on, we assume that Eq. \ref{eq:forb} holds, in other words, we assume that Eq. \ref{eq:mpi} has a solution. In this case, any vector $\textbf{q}$ of the form
\begin{equation}
\label{eq:06}
\textbf{q}=\textbf{B}_p^{+}\textbf{b}_p +({\textbf{Id}-\textbf{B}_p^+ \textbf{B}_p})\nu
\end{equation}
 solves the linear equation system Eq. \ref{eq:mpi} where $\textbf{Id}$ is the identity matrix and $\nu$ is any column vector of the right dimension. In fact, these are all the solutions to Eq. \ref{eq:mpi}. Summarizing the above discussion, we have at least one solution to Eq. \ref{eq:mpi} if and only if Eq. \ref{eq:forb} holds for $\textbf{b}_p$. Moreover, if there is a solution to Eq. \ref{eq:mpi}, then all solutions have the form of Eq. \ref{eq:06} \cite{Moore1920,Penrose1955}.
 
 
In Eq. \ref{eq:06}, the parameter $\nu$ is freely chosen by the user to control the solution. In our examples, we take $\nu$ to be the initial edge lengths of the polyhedron, resulting in a solution to Eq. \ref{eq:06} which is the new geometry for the initial polyhedron with the prescribed area for face $f_i$. In this case, the new edge lengths are the best fit (least squares) to the initial edge lengths. Also, in many cases, only certain parts of the polyhedron change significantly (see Figure \ref{fig:face_comp} and \ref{fig:app1}). 

Another approach could be to take $\nu$ to be the vector whose entries are all 1, in this case, we get a solution with well-distributed edge lengths.

\subsubsection{Updating the Polyhedral geometry with multiple prescribed face areas}
The previously discussed method can compute the geometry of a polyhedral system with multiple faces with prescribed areas in an inductive process. In each step, we update the geometry of the polyhedron using Eq. \ref{eq:06} with the newly computed face whose edge lengths are added to the list of fixed edges. The new edge lengths change the constraints equations $\mathbf{L}_p\mathbf{q}=\textbf{l}_p$ to compute the polyhedral geometry.

We summarize the process in Algorithm \ref{alg:poly}.

%% file: method_05.tex
\subsection{Updating the internal forces in the dual diagram}\label{sec:dual}
\begin{figure}
 \centering
  \includegraphics[width=\columnwidth]{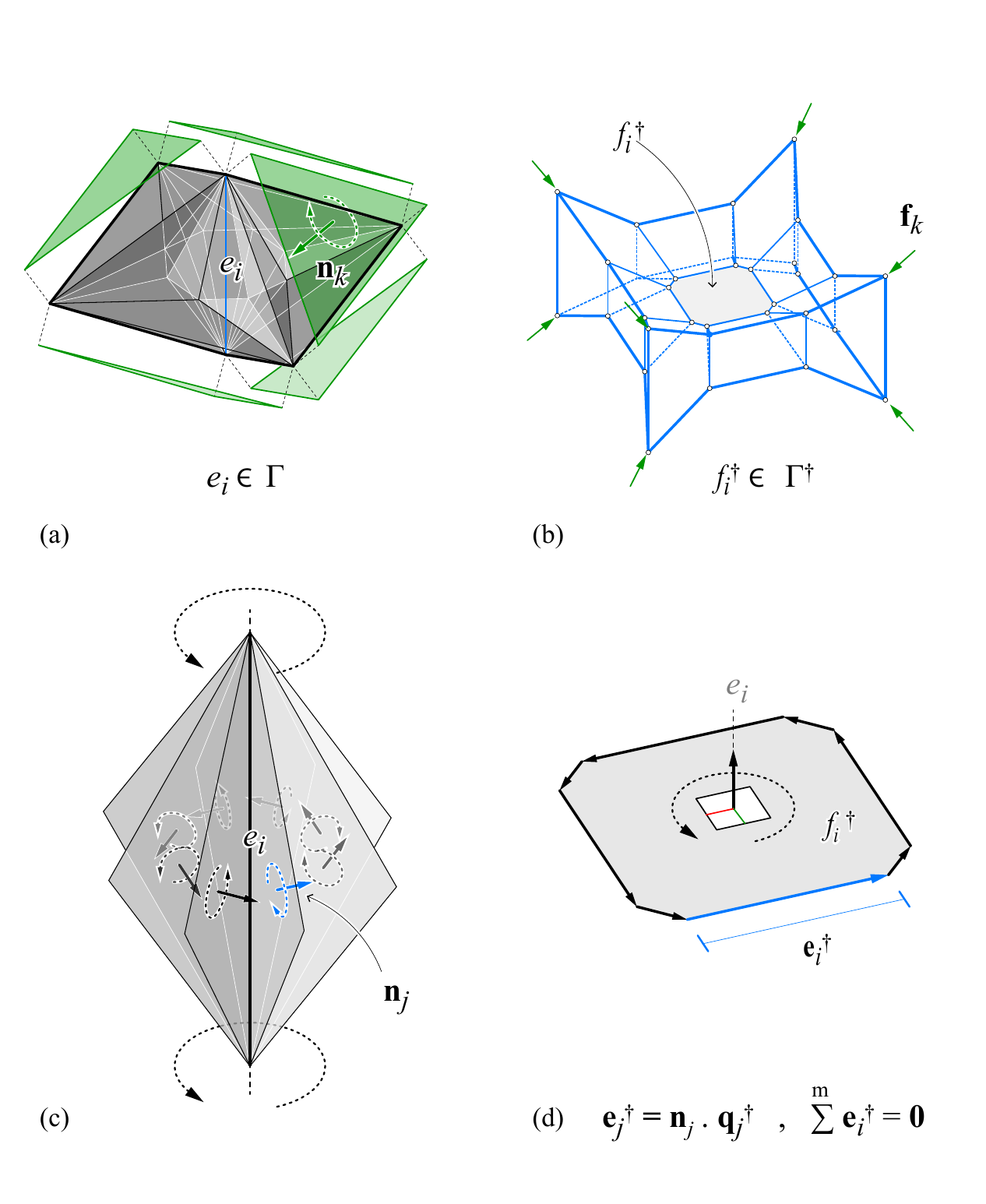}
  \caption{ (a) An input force polyhedron as primal and its corresponding (b) funicular polyhedron as the dual; (c) going around each edge of the primal with its attached faces (c) provides the direction of the edge vectors of the corresponding face (e) in the reciprocal diagram where the sum of the edge vectors must be zero.
  }
  \label{fig:dof1}
\end{figure}

In \cite{akbarzadeh2018developing, HABLICSEK201930} three different methods were described to generate the dual diagram from a given primal. 
 Let us call the starting diagram the \textit{primal}, $\Gamma$, and the reciprocal perpendicular polyhedron \textit{dual}, $\Gamma^\dagger$ (Fig. \ref{fig:dof1}a, b). The vertices, edges, faces, and cells of the primal are denoted by $v$, $e$, $f$, and $c$ respectively, and the ones of the dual are super-scripted with a dagger ($\dagger$) symbol (Figure \ref{fig:dof1}a,b).




Since the face ${f_i}^\dagger$ is a closed polygon, the sum of the edge vectors ${\textbf{e}_j}^\dagger$ should be zero. Hence, we obtain a vector equation similar to Equation \ref{eq:eqsysface}
\[\sum_{f_j} {\textbf{u}_j}^\dagger q_j^\dagger=\textbf{0}\]
where the sum runs over the attached faces $f_j$ of the edge $e_i$ of the primal $\Gamma$; $\textbf{u}_j^\dagger$ denotes the unit directional vector corresponding to the edge vector $\textbf{e}_j^\dagger$; and $q_j^\dagger$ denotes the edge length of $\textbf{e}_j^\dagger$ in the dual $\Gamma^\dagger$.



Similarly, as before, each vector equation for the face of the dual diagram yields three linear equations for the edge lengths, and we obtain a linear equation system for the edge length vector $\textbf{q}^\dagger$ which can be described by a $[3e\times f]$ matrix that we call the equilibrium matrix of the dual diagram, $\textbf{E}^\dagger$,
\begin{equation}\label{eq:04}
    \textbf{E}^\dagger\textbf{q}^\dagger=\textbf{0}.
\end{equation}




In \cite{akbarzadeh2018developing, HABLICSEK201930} three different methods were described to generate the dual diagram from Eq. \ref{eq:04}. In this paper, we choose to use the Moore-Penrose Inverse (MPI) method to initially construct the dual diagram before we apply any changes to the force and its face areas. The MPI method of this section is as same as the method described in Section \ref{sec:mpi} with $\mathbf{b}_p$ being the zero vector. As a result, the solutions to Eq. \ref{eq:04} can be described as
\[\textbf{q}^\dagger=\left(\textbf{Id}-(\textbf{E}^{\dagger})^{+} \textbf{E}^\dagger\right)\xi.\]
Here $(\textbf{E}^{\dagger})^+$ denotes the MPI of the equilibrium matrix $\textbf{E}^\dagger$. For the parameter $\xi$,  we choose the vector whose entries are all 1 to obtain a dual diagram with well-distributed edge lengths. 

\subsubsection*{The direction of the internal forces}

The initial direction of the internal force as compression or tension is stored and altered after the computation of the force with prescribed areas. The tensile force members are updated in the form if the normal of a face in the force diagram flips after the computation. As shown, the geometry of a face can become self-intersecting in some cases. On such occasions, if the area of the region with the initial normal direction is bigger than the area with the flipped normal, then the direction of the initial internal force does not change; otherwise, the direction of the internal force will flip. If the face is a zero area face, the member will carry no force and can disappear in the form diagram (Figure \ref{fig:app1}.)

%% file: implementation.tex
\section{Implementation}

\begin{figure}
    \centering
    \includegraphics[height = .9\textheight]{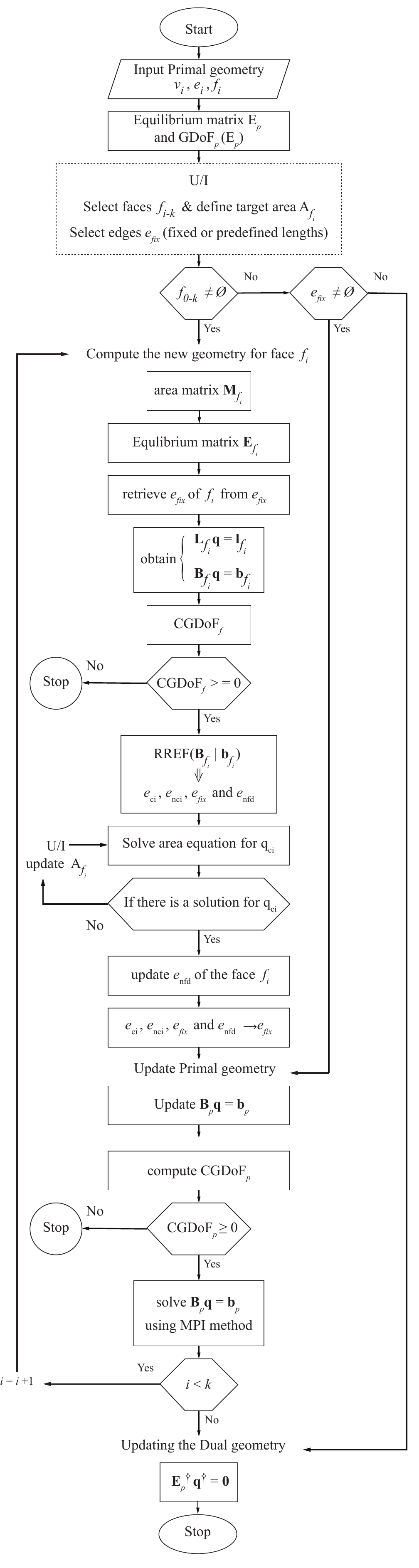}
    \caption{The flowchart expanding multiple steps in computing the primal geometry/force diagram with preassigned edge lengths and face areas, and the updated dual geometry as the form diagram.}
    \label{fig:flowchart}
\end{figure}

In this section, we explain the computational setup for the methods we explained in Section \ref{sec:meth}. The input for this framework is a polyhedral system with planar faces and is considered as the force diagram for the methods of 3D graphic statics. we compute the dual geometry of the updated primal diagram according to \cite{akbarzadeh2018developing, HABLICSEK201930}.

The user can initially select certain edges in the system and assign a target length per selected edge. Similarly, s/he can select multiple faces and assign an area per face. Our method is a sequential computational setup that updates the geometry of the polyhedral system for each user-selected face at each step. For instance, let's assume that the user selects three faces with a target area per face. We start from a face, compute the new geometry of the face with the target area, update the geometry of the polyhedral system based on the new geometry of the face, and move on to the next face and continue the computation until there is no face left.     

Accordingly, in each step of the computation, we construct the area matrix and the equilibrium matrix for the user-selected face. We, then, add the user-defined edge constraints as non-homogeneous linear equations and compute the Constrained Geometric Degrees of Freedom, $CDGoF$ using RREF method explained in Section \ref{sec:face}. In the next step, we compute the geometry of the face, and then we update the geometry of the polyhedral system using MPI method described in Section \ref{sec:poly}.

After the multi-step computation process is completed, we update the direction and the magnitude of the forces in the members of the dual that was initially constructed. 



The above description can be summarized into three main sections as shown in the flowchart of Figure \ref{fig:flowchart}. These sections are as follows: 

\begin{itemize}
    \item computing the new geometry for a face with constrained edges and areas;
    \item updating the new geometry of the polyhedral system based on the newly-computed face geometry and the fixed edges; and, 
    \item updating the internal forces in the members of the dual based on the new force magnitudes.
\end{itemize}

Sections \ref{sec:impa}, \ref{sec:impf}, and \ref{sec:impu} provide additional details of the algorithms used in this process. These algorithms include: computing the area matrix for a face; face reconstruction with constrained edges and target area; and updating the geometry of the polyhedron with constrained areas of faces and edge lengths.




\begin{algorithm}
\DontPrintSemicolon
\KwIn{
$f_i: [v_{0}, v_n,..., v_{k-1}]$ the ordered list of vertices around $f_i$
}
 \KwOut{$\mathrm{\textbf{M}}_{f_i}$ the area matrix of the face $f_i$.}
 \SetKwFunction{FMain}{$M_{row}$}
 \SetKwProg{Fn}{Function}{:}{}
 \Fn{\FMain{$\overline{v}_{row}$, ${\mbox{\textbf{n}}}_{f}$}}{
$\overline{v}_{row}: [v_j, v_{j+1},..., v_{j-1}]$ {\color{gray}{\# ordered list of vertices starting from $v_j$}}\;

\For {$v_n \in \overline{v}_{row}$}{
$\overline{e} \longleftarrow e_i[v_n, v_{n+1}]$ {\color{gray}{\# ordered edges }}\;  
} 
  
\For{$e_i \in \overline{e}$}{
$ \textbf{e}_{i} \longleftarrow \langle x_{n+1}-x_n, y_{n+1}-y_n, z_{n+1}-z_n \rangle$ {\color{gray}{\# direction vector from $v_n$ to $v_{n+1}$}} \;
$ |\textbf{e}_{i}| \longleftarrow {l_i}$ {\color{gray}{\# length of the vector $\textbf{e}_{i}$}}\;
}

\For{$e_i \in E$}{
$\textbf{n}_{0, i}  = \textbf{e}_0 \times \textbf{e}_{i}  $ {\color{gray}{\# cross product of $\textbf{e}_0$ and $\textbf{e}_{i}$}}\;
\eIf{$x_{\textbf{n}_f} \neq 0$ {\color{gray}{\# the $x$ coordinate of $\textbf{n}_f$}}}
    {$\eta_{i} = x_{\mbox{n}_{0, i}}/(x_{\textbf{n}_f} * |\textbf{e}_{i}| * |\textbf{e}_0|)$ \;}
    {
    \If{$y_{\textbf{n}_f} \neq 0$ {\color{gray}{\# the $y$ coordinate of $\textbf{n}_f$}}}
    {$\eta_{i} = y_{\textbf{n}_{0, i}}/(y_{\textbf{n}_f} * |\textbf{e}_{i}| * |\textbf{e}_{0}|)$ \;}
    \If{$z_{\textbf{n}_f} \neq 0$ {\color{gray}{\# the $z$ coordinate of $\textbf{n}_f$}}}
    {$\eta_{i} = z_{\textbf{n}_{0, i}}/(z_{\textbf{n}_f} * |\textbf{e}_{i}| * |\textbf{e}_{0}|)$ \;}
    }
    $M_{e_i} = (k - 1 * i - 1) * \eta_i$ {\color{gray}{\# matrix coefficient for edge $e_i$ where $k$ is the length of $V_{row}$ }}\;
    $\textbf{M}_{row} \longleftarrow M_{e_i}$\;
}
  
 \KwRet  $\textbf{M}_{row}$\;
  }

\Begin{
$\textbf{e}_0 \longleftarrow$ {\color{gray}{\# vector from $v_0$ to $v_1$}}\;
$\textbf{n}_{f_i} \longleftarrow \textbf{e}_0 \times \textbf{e}_{1}$ {\color{gray}{\# the cross product of the first two edges of $f_i$}}\; 
\For{$v_i \in \{v_0,...,v_{k-1}\}$}{
$\overline{v}_{v_i} \longleftarrow [v_i, v_{i+1}, ..., v_{i-1}]$ {\color{gray}{\# ordered list of vertices starting at $v_i$}}\;
$\textbf{M}_{v_i} = \textbf{M}_{row}(\overline{v}_{v_i}, \mbox{\textbf{n}}_{f_i})$
}
$\textbf{M}_{f_i}\longleftarrow \textbf{M}_{v_i}$ {\color{gray}{\# matrix whose rows are the $\textbf{M}_{v_i}$}}\;

$\textbf{M}_{f_i}=\frac{1}{2}\left(\textbf{M}_{f_i}+\textbf{M}_{f_i}^T\right)$ {\color{gray}{\# output is a symmetric matrix}}\;

}
\caption{Computing the area matrix $\textbf{M}_{f_i}$}
\label{alg:area}
\end{algorithm}

\subsection{Constructing the area matrix}\label{sec:impa}
Algorithm \ref{alg:area} receives a face $f_i$ with an ordered list of vertices $[v_0,...,v_{k-1}]$ as an input and outputs a symmetric matrix $\textbf{M}_{f_i}$ which is used in the quadratic form, Eq. \ref{eq:quadrform}, to compute the area of the face.

First, we choose an (arbitrary) normal vector for the face, by taking the unit cross product of the first two consecutive edge vectors. Then, we construct the matrix $\textbf{M}_{f_i}$ row by row as follows: starting from each vertex $v_n$, we create an ordered list of directed edges $\textbf{e}_i$, and compute the scalars $\eta_{i,j}$ as explained in Eq. \ref{eq:eta}.

Once the whole matrix $\textbf{M}_{f_i}$ is constructed, the algorithm outputs a symmetric matrix (see Eq. \ref{eq:symmat}) to be used in the quadratic form for computing the area of the face $f_i$.

\subsection{Updating the geometry of a face with constrained edges and a target area}\label{sec:impf}

The Algorithm \ref{alg:AF} updates the geometry of a face with constrained edges and a target area. The input of this algorithm is a user-selected face $f_i$ with a target area $A_{f_i}$ and a (user-selected) edges of $\overline{e}_{fix}$ with prescribed edge lengths. 

First, we compute the linear constraint equations, Eq. \ref{eq:cons}, where the equilibrium equations describe the geometry of the face and the constraint equations come from the (user-selected) edge constraints. 

Once, the constraint equation system, Eq. \ref{eq:cons}, is created, we compute the Constrained Geometric Degrees of Freedom of $f_i$, $CDGoF_{f_i}$, of the face using RREF method. If $CDGoF_{f_i}=-\infty$, the algorithm stops, i.e., the constrained equation system, Eq. \ref{eq:cons}, cannot be solved. In this case, the user may modify the input by selecting less constrained edges or by selecting different edges.

If $CDGoF_{f_i}$ is at least zero, the algorithm classifies the edges of $f_i$ into \textit{ci},\textit{ nci}, \textit{fix} and \textit{nfd} edges. Next, the we construct the area matrix $\textbf{M}_{f_i}$ using Algorithm \ref{alg:area}. Using the area equation, Eq. \ref{eq:quadrform}, and Eq. \ref{eq:quadr}, we compute the edge length $q_{ci}$ of the edge $e_{ci}$. If Eq. \ref{eq:quadr} has no solution, the algorithm may ask the user to modify the target area. The Eq. \ref{eq:quadr} often has two solutions, the user may choose from those particular solutions (see Figure \ref{fig:two}).

Once a solution is chosen for $q_{ci}$, the algorithm updates the lengths of the $e_{nfd}$ (see Section \ref{sec:updatefi}) and outputs the updated lengths of the edges of $f_i$.

\begin{algorithm}
\DontPrintSemicolon
\KwIn{$\begin{cases}
f_i: [v_0,...,v_{k-1}] & $face with ordered list of vertices$\\ 
A_{f_i} & $target area for the face$ \\ 
\overline{e}_{fix}:[e_m,...,e_q] & $list of constrained edges$\\
\end{cases}$
}
\KwOut{$Q:[q_{0},...,q_{k-1}]$ list of edge lengths of $f_i$ with area $A_{f_i}$.}

%

\Begin{
$\overline{e} \longleftarrow [e_0, ..., e_{k-1}]$ {\color{gray}{\# ordered list of edges}}\;

\For{$e_i \in \overline{e}$}{
$ \textbf{e}_{i} \longleftarrow \langle x_{n+1}-x_n, y_{n+1}-y_n, z_{n+1}-z_n \rangle$ {\color{gray}{\# direction vector from $v_n$ to $v_{n+1}$}} \; 
$ |\textbf{e}_{i}| \longleftarrow {l_i}$ {\color{gray}{\# length of the vector $\textbf{e}_{i}$}}\;
$\textbf{u}_i=\textbf{e}_i/|\textbf{e}_i|$ {\color{gray}{\# unit direction vector of $e_i$}}\;

$\textbf{E}_x \longleftarrow \textbf{x}_{\textbf{u}_i}$ {\color{gray}{\# row vector of the $x$-coordinates of $\textbf{u}_i$}}\; 
$\textbf{E}_y \longleftarrow \textbf{y}_{\textbf{u}_i}$ {\color{gray}{\# row vector of the $y$-coordinates of $\textbf{u}_i$}}\; 
}
$\textbf{E}_{f_i} \longleftarrow \textbf{E}_x, \textbf{E}_y$ {\color{gray}{\#  the $[2\times e]$ equilibrium matrix of the face $f_i$}}\;  

\For{$e_i \in \overline{e}_{fix}$}{
$\textbf{l}_i, q_i$ {\color{gray}{\# the row vector of the constraint equation for a fixed edge Eq. \ref{eq:eqlf}}}\;
}
$\textbf{B}_{f_i}, \textbf{b}_{f_i}$ {\color{gray}{\# create the constraint equation system Eq. \ref{eq:cons}}}\;

$RREF((\textbf{B}_{f_i}|\textbf{b}_{f}))$ {\color{gray}{\# compute the RREF of the constraint equation system}}\;
$CGDoF_{f_i}$ {\color{gray}{\# compute the CGDoF of the system}}

\If{$CGDoF_{f_i}=-\infty$}{
no solution $\Longrightarrow$ end program or ask user to modify the input.\;
}
\Else{$ \textbf{q}_{nci}$ {\color{gray}{\# identify the nci edges}}\;
$\textbf{D}, \textbf{d}, \textbf{g}$ {\color{gray}{\# coefficients of Eq. \ref{eq:q_dep}}}\;
}
$\textbf{D}', \textbf{d}'\longleftarrow \textbf{D}, \textbf{d}$ {\color{gray}{\# coefficients of Eq. \ref{eq:qform}}}\;
$\textbf{M}_{f_i}$: {\color{gray}{\# output of Algorithm \ref{alg:area}}}\; 
$a, b, c$ {\color{gray}{\# coefficients of Eq. \ref{eq:quadr} using $\textbf{M}_{f_i}$}}\;
$q_{ci}\longleftarrow a, b, c$ {\color{gray}{\# compute the solution(s) of Eq. \ref{eq:quadr}}}\;
$\textbf{q}_{nfd}\longleftarrow q_{ci}, \textbf{q}_{nci}, \textbf{q}_{fix}$ {\color{gray}{\# update the edge lengths}}\;
$Q=[q_0,...,q_{k-1}]$ {\color{gray}{\# the list of updated edge lengths}}\;

}
 \caption{Updating the geometry of the face (UF)}
 \label{alg:AF}
\end{algorithm}

\subsection{Updating the geometry of the polyhedron}\label{sec:impu}
The last algorithm updates the geometry of a polyhedron after the new geometry of each face is computed. In fact, this is a multi-step process, in each step, we update the geometry of the polyhedron after updating the geometry of a face. Note that, in each step, the list of fixed edges is updated after computing the geometry of each face, in other words, we solve for the geometry of the polyhedron with more and more constraints.

The Algorithm \ref{alg:poly} describes the computation of the new geometry of a polyhedron with a given list of constrained edges, $\overline{e}_{fix}$. It computes the linear constraint equation system, Eq. \ref{eq:cons2}. This is a non-homogeneous linear equation system which can be solved using MPI method. The parameter $\nu$ in the MPI method can be chosen by the user or can be the vector of the initial values of the edge lengths of the polyhedron.

\begin{algorithm}[h!]
\DontPrintSemicolon
\KwIn{
$\overline{e}_{fix}:[e_{m}, e_n,...,e_q]:$  list of fixed edges\\
}
\KwOut{$Q_p:[q_{1},...,q_e]$ the updated list of edge lengths of the polyhedron}

\Begin{
\For{$e_i\in \overline{e}_{fix}$}{
$\textbf{l}_i, q_i$ {\color{gray}{\# the row vector of the constraint equation for a fixed edge (Eq. \ref{eq:eqlf})}}\;
}
$\textbf{B}_{p}, \textbf{b}_p$ {\color{gray}{\# create the constraint equation system (Eq. \ref{eq:cons2})}}\;
$\textbf{B}_p^+\longleftarrow$MPI$(\textbf{B}_p)$\; 
$Q_p$ {\color{gray}{\# update the edge lengths of the polyhedron using MPI method with a fixed parameter $\nu$ (Eq. \ref{eq:cons2})}}
}

 \caption{Updating the geometry of the polyhedron}
 \label{alg:poly}
\end{algorithm}

%% file: application.tex
\begin{figure*}
    \centering
    \includegraphics[width = \columnwidth]{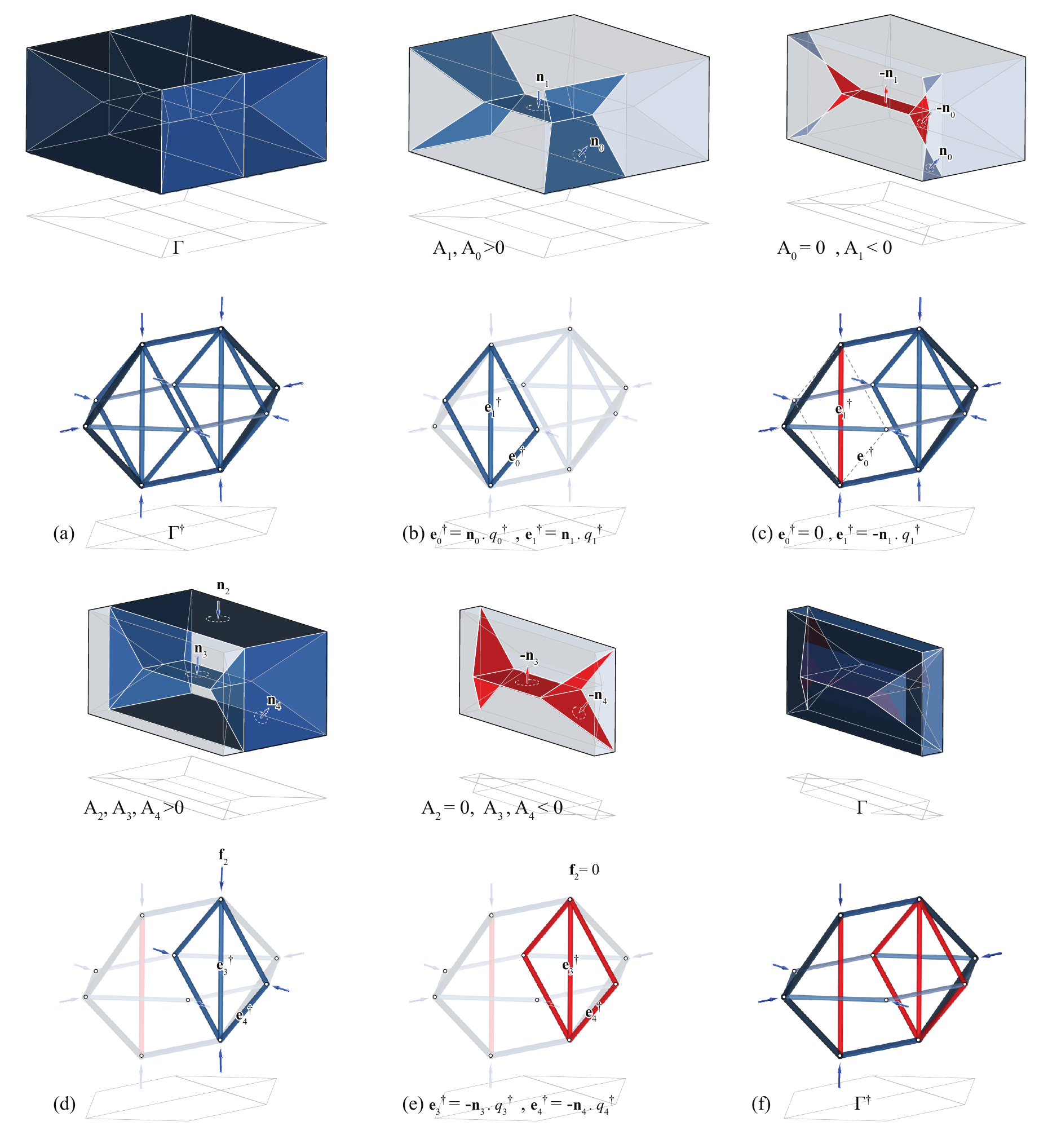}
    \caption{(a) Initial force polyhedron of the Figure \ref{fig:face_comp}j (top) and its reciprocal form (bottom) as a compression-only system; (b) highlighted internal faces of the polyhedron before transformation (top) and the corresponding members in the form (bottom) ; (c) zero area faces and their updated normal directions and the resulting zero-force members in the form; (d), (e) and (f) highlighted faces in the second step of the transformation and the updated form with new internal force distribution.}
    \label{fig:app1}
\end{figure*}

\section{Applications/results}\label{sec:app}

\begin{figure*}
    \centering
    \includegraphics[width = \textwidth]{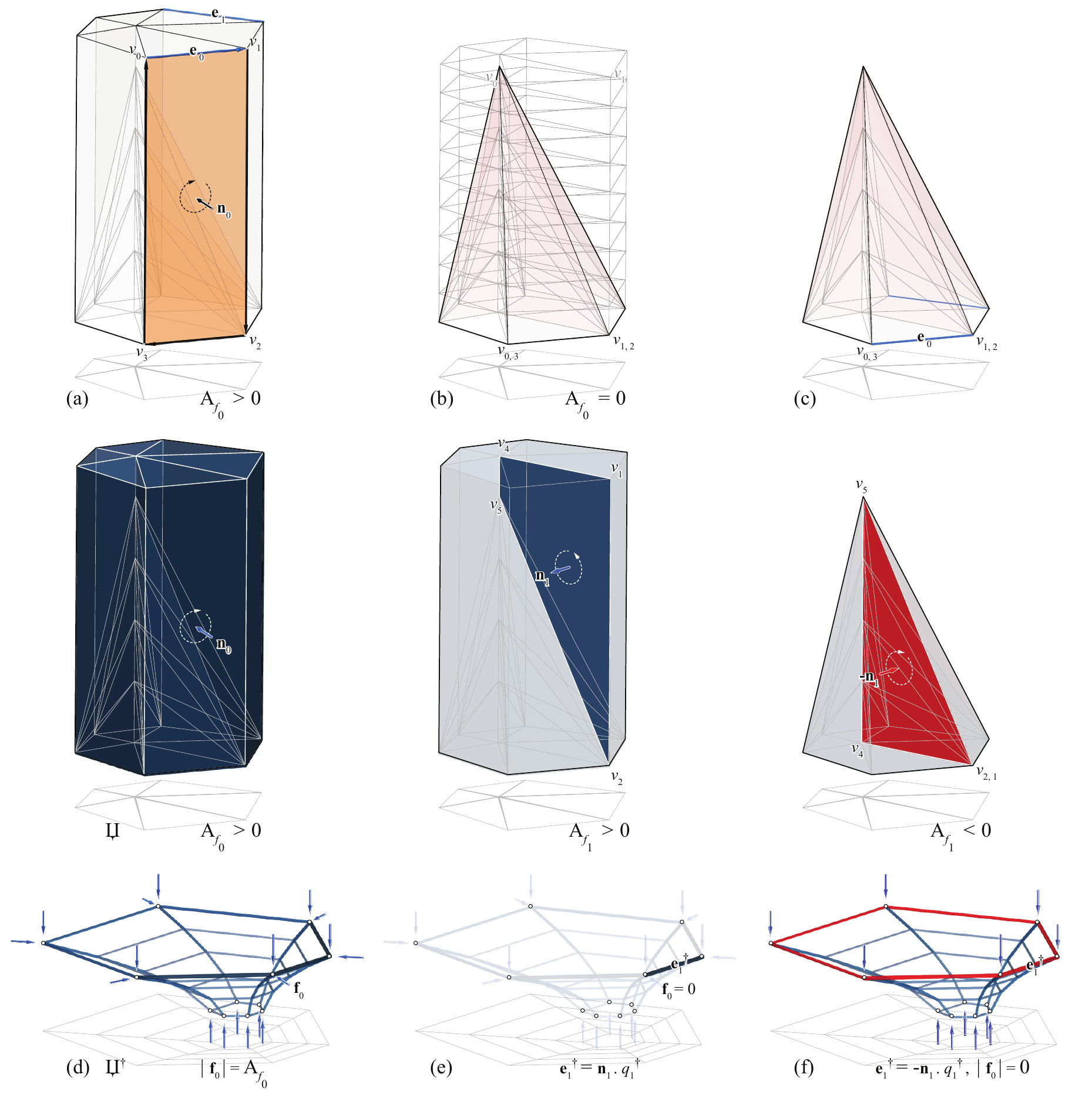}
    \caption{(a) A force polyhedron and a selected face with zero area target, and two fixed edges on the top; (b) the transformation animation (for visualization purposes only); (c) the resulting force polyhedron where the selected face and all other side faces collapse to a line; (d), (e), and (f) the force and form diagrams and their transformation after changing the areas.}
    \label{fig:app2}
\end{figure*}

Figures \ref{fig:face_comp}, \ref{fig:app1}, and \ref{fig:app2}   illustrate the potentials of using this approach in polyhedral transformation with target areas of faces and constrained (selected) edge lengths, and their corresponding structural forms. In the examples shown in Figure \ref{fig:face_comp}, the intention is to highlight certain properties of the constrained polyhedral computation. In all the examples the chosen face is highlighted by an orange shade and the constrained edges are highlighted by blue color.

For instance, in Figure \ref{fig:face_comp}a, the target area for the chosen face is zero and an edge has been chosen that does not belong to the selected face. Figure \ref{fig:face_comp}b shows an animated drawing for clarification purposes. As it is shown in Figure \ref{fig:face_comp}c, the selected face turns into a self-intersecting face. In the next example, in Figure \ref{fig:face_comp}d, the top face is set to be zero with the same constrained edge. As a result of this computation, the face collapses to a line due to its rectangular geometry. Also, the normal of the side faces, $\textbf{n}_{f_{0}}$, flips which means that the magnitude of the internal force in the corresponding form will change (\ref{fig:face_comp}h, i). In the Example of Figure \ref{fig:face_comp}g, one of the vertical side faces with a constrained top edge is chosen and the target area is also set to zero. As illustrated in Figures \ref{fig:face_comp}h and \ref{fig:face_comp}i, the rest of the vertical faces will disappear after the polyhedral computation. This method can certainly be applied to more than one face in the polyhedral system. In Figure \ref{fig:face_comp}j, one external and one internal face are chosen with a zero area target and as illustrated in Figures \ref{fig:face_comp}k-o, the computation process proceeds sequentially by first, computing the geometry of the face $f_0$ in Figure \ref{fig:face_comp}l and then recomputing the polyhedral geometry to solve for the new face $f_1$. In this process, multiple other faces will also turn into a self-intersecting face as shown in Figure \ref{fig:face_comp}l. In the final geometry, the face $f_1$ will collapse to an edge $e_1$ (Figure \ref{fig:face_comp}o).

In most of these examples, the target areas were intentionally set to be zero to highlight its results in the reciprocal structural form.  
Figure \ref{fig:app1}a shows the same polyhedron of the Figure \ref{fig:face_comp}j. In this figure, certain faces in the force diagram and their corresponding edges in the form diagram are highlighted to show the effect of changing the areas on the magnitude of the internal forces. Starting from the force polyhedron of Figure \ref{fig:app1}a (top) and its compression-only form (bottom), faces $f_1$ and $f_0$ are emphasized in Figure \ref{fig:app1}b (top) and their corresponding compression-only edges in the form. The result of the zero area computation results in face $f_0$ to turn into a self-intersecting face together with other similar faces attached to the edges of the face $f_1$. Besides, face $f_1$ is flipped as well as the direction of its normal vector. Note that as a result of this transformation, the internal force in the corresponding member of the face $f_0$ is decreased to zero. This is a fascinating effect in the equilibrium of polyhedral frames as it describes the internal equilibrium of forces in a polyhedral system where the edges or the members can be removed from the system without disturbing the internal and external equilibrium. The zero-force edges have been previously observed in some polyhedral reciprocal diagrams by \cite{McRobie2016}, but there was no method to compute them particularly in self-intersecting faces as described in the methodology section. Also, the change in the direction of the normal of the face $f_1$ results in reversing the internal force in the edge $e_1$ of the form diagram. 

Figure \ref{fig:app1}d emphasizes the faces $f_2$, $f_3$, and $f_4$ and faces that will be affected as a result of the second step of the computation. As it is shown in Figure \ref{fig:app1}e, the normal of the faces $f_3$ and $f_4$ will invert the direction of the internal forces in the form diagram. This transformation also removes the applied load in the system as the area of the face $f_2$ is zero. In another example, the zero area faces are used to completely remove the external horizontal forces in the system. Figure \ref{fig:app2} shows a force diagram and its corresponding form in another transformation. In this example, one of the vertical faces is chosen and the target area is set to become zero. Note that as animated in the drawings of Figure \ref{fig:app2}b, all the side faces collapse into a line and the top and bottom faces of the polyhedron will become coplanar. This transformation results in the disappearance of all the horizontal applied loads in the system as shown in Figure \ref{fig:app2}f. The most interesting geometric outcome of this process is the transformation of the internal face $f_1$ in this process. Face $f_1$ changes its direction and so does its corresponding edge at the boundary of the form diagram. The resulting structure shows a funnel shape compression-only structure with tensile members on the top.

%% file: conclusion.tex
\section{Conclusion and discussion}
This paper provides an algebraic formulation alongside with algorithms, and numerical methods to geometrically control the areas of the faces of general polyhedrons of the reciprocal diagrams of 3D/polyhedral graphic statics. The presented methods bridge the gap between the previously developed algebraic methods for the construction of the reciprocal polyhedral diagrams and controlling the magnitude of internal and external forces by changing the areas of the faces. This method for the first time allows the user to manipulate both convex and complex faces and explore the compression and tension combined features in structural form-finding using 3DGS. Controlling the areas of complex faces has never been addressed in the literature prior to this research as the previous approaches mainly dealt with convex polyhedrons. Thus, this research opens a new horizon understanding the equilibrium of both tension and compression forces beyond the existing compression-only polyhedral funicular forms. 

The paper explains the process of turning geometric constraints such as edge lengths and target face areas of the reciprocal polyhedral diagrams into algebraic formulations compatible with the previously developed method by \cite{HABLICSEK201930} and \cite{akbarzadeh2018developing}. This research describes a quadratic formulation to compute the geometry of a face with a target area and provides a linear formulation to consider the edge lengths as constraints. Solving an equation system including both linear and quadratic equations is a highly complex task. The key idea in our proposed method is to reduce the number of unknowns in the (quadratic) equation system of a face using Reduced Row Echelon (RREF) method. Computing the updated geometry of the polyhedral diagrams is achieved by Moore-Penrose Inverse (MPI) method. In this approach, multiple faces and edges can be selected as constraints, and the new geometry of the polyhedral system is computed in a sequential process.  

The paper also describes the Constrained Geometric Degrees of Freedom (CGDoF) of the linearly constrained polyhedral systems and opens up a door to explore a wide variety of interesting geometries satisfying the initial equilibrium equations with selected edge lengths. The algorithms and numerical methods provide an interactive tool for the user to study and manipulate large-scale general polyhedral diagrams by assigning face areas and edge lengths.

\subsection*{Future work}
The existing algorithm deals with each face at each step and adds the newly computed edge lengths to the initial constraints of the polyhedral system. As a consequence there is no control over the number of constraints generated in each step and the eventual number of constraints to compute the entire polyhedral group. Therefore, in certain cases, depending on the chosen edge by the user, or the geometric degrees of freedom of the entire system, the polyhedral computation becomes over-constrained. This property proposes an interesting problem for future research.

Another interesting future direction is the study of the different solutions of the same initial, constrained problem. As mentioned in Section \ref{sec:number}, for a single face there can be two, significantly different polygons satisfying the linear and quadratic constraint equations. As a consequence, for a polyhedral system with $n$ assigned face areas, there can be $2^n$ significantly different updated polyhedral systems that can also be explored in future research. 

\section{Acknowledgment}

This research was partially funded by National Science Foundation Award (NSF CAREER-1944691). 